 \documentclass[final,3p,times,number]{elsarticle}

\usepackage{amssymb}
\usepackage{lipsum}
\usepackage{amsmath}
\usepackage{booktabs}
\usepackage{float}
\usepackage{graphicx}
\usepackage{subcaption}
\usepackage{upgreek}
\usepackage{placeins}
\usepackage[font=normalsize, labelfont=bf]{caption}
\usepackage[font=normalsize]{subcaption}
\usepackage[labelfont=normalfont]{subcaption}

\usepackage{xcolor}

\journal{Elsevier}

\begin{document}

\begin{frontmatter}



\title{An efficient and energy stable framework for phase field simulations of grain growth in additive manufacturing}


\author[first]{Chaoqian Yuan}
\author[second,third]{Chinnapat Panwisawas}
\author[first]{Ye Lu\corref{cor1}}
\ead{yelu@umbc.edu}
\affiliation[first]{organization={Department of Mechanical Engineering, University of Maryland Baltimore County},
            city={Baltimore},
            country={USA}}
\affiliation[second]{organization={School of Engineering and Materials Science, Queen Mary University of London},
            city={London E1 4NS},
            country={UK}}
\affiliation[third]{organization={Metallurgy and Materials Science Research Institute, Chulalongkorn University},
            city={Bangkok 10330},
            country={Thailand}}
 \cortext[cor1]{Corresponding author}

\begin{abstract}
Phase field simulations play a key role in the understanding of microstructure evolution in additive manufacturing. However, they have been found extremely computationally expensive. One of the reasons is the small time step requirement to resolve the complex microstructure evolution during the rapid solidification process. This paper investigates the possibility of using a class of stabilized time integration algorithms to accelerate such phase field simulations by increasing the time steps, based on a phase field model dedicated to simulating the solidification of 316L stainless steel during additive manufacturing, particularly in a regime where the solid-liquid interface is moving fast and there is absolute interfacial stability with negligible
composition variations. The specific computational framework, incorporating the finite element method and the stabilized time integration algorithms, was developed. A theoretical analysis on energy stability was conducted, based on a revisited  energy law derived for the phase field model. The numerical results confirmed that the proposed framework can effectively enforce the numerical stability and a decreasing energy requirement for the phase field simulations with at least two orders-of-magnitude larger time steps over conventional explicit methods. 2D and 3D phase field simulations have been conducted with relevant physical and kinetic parameters for 316L stainless steel.  This computational framework can be easily adapted for different phase field models and open numerous opportunities for efficient phase field simulations.
\end{abstract}


\end{frontmatter}


\section{Introduction}
\label{introduction}
Additive manufacturing (AM) is a promising manufacturing technology for producing complex parts with high design flexibility and minimal material waste \cite{Frazier2014,Gibson2015,Herzog2016review,Wang2018hierarchical_stainless,Bajaj2020,Armstrong2022}.  With appropriate post-processing treatments, 3D printed parts have been shown to match or even surpass the performance of conventionally manufactured counterparts \cite{DebRoy2018}. However, the complex thermal cycles, including the rapid heating and cooling rates, involved in the printing processes (e.g., laser powder bed fusion) often lead to complex heterogeneous microstructures in the printed materials \cite{Song2015}. In addition, numerous studies have shown that process parameters, including heat source power, scan speed, and scan strategy, have a significant influence on microstructure evolution and the resulting mechanical properties \cite{Wang2016,Kok2018, Kurose2020, Guo2020, Li2020, Henry2021}. Therefore, understanding the microstructure evolution and their relationship to process parameters is crucial for improving the quality and performance of parts produced by AM processes.

Numerical simulations have been developed to understand the microstructure evolution in AM processes. They are usually based on numerical models developed for solidification processes \cite{Kurz2020}.  Three approaches can be used, including cellular automata (CA) \cite{Gandin1999, Zinovieva2018, Lian2019, Yu2021, Xie2023}, kinetic Monte Carlo (KMC) method \cite{Battaile2008, Rodgers2017, Rodgers2018, Johnson2018, Rodgers2021} and phase field method (PFM) \cite{chen2002phase, Boettinger2002, Keller2017, ji2018understanding, Dubrov2019, MY2021, Chadwick2021, Chadwick2025}. Compared to PFM, the primary advantage of CA and KMC methods lies in their relatively low computational cost, which enables large volume simulations for understanding the overall texture development in 3D printed materials. They can capture several key features of AM microstructure, like the epitaxial growth of columnar grains. However, their rule-based framework may limit the generalizability. For example, the traditional CA model developed by Gandin and Rappaz employs stochastic rules to represent dendritic growth \cite{Rappaz1993,Gandin1994,Gandin1999}, which might be difficult to capture the planar transition of the solid–liquid interface during the increase of interfacial velocity \cite{Merchant1990}. In the KMC approach \cite{Rodgers2017, Rodgers2018, Rodgers2021}, due to the lack of continuous field representation, the coupling with complex thermal and mechanical fields may be difficult. The PFM is regarded as a more accurate approach for simulating microstructure evolution, as it is derived from irreversible thermodynamics and provides a natural framework for incorporating driving forces of different nature. Nevertheless, due to the high computational expenses of the PFM, the simulations are often limited to 2D cases or small volumes with simplified physical assumptions \cite{Lu2018, Liu2018, Liu2019ColEQT, MY2021,Chadwick2021}. Therefore, accelerating phase field simulations is crucial to enlarge the computational domain, incorporate  multiphysics coupling effects, and validate the numerical models with experimental samples. 

Various techniques have been developed to accelerate phase field simulations by reducing the computational complexity in the PFM or constructing surrogate models. These include adaptive meshing \cite{provatas1998efficient,zhu2019phase,dewitt2020prisms}, high order approximation methods \cite{feng2006spectral,li2024adaptive}, data and physics-informed machine learning \cite{liu2019multi,montes2021accelerating,peivaste2022machine,xue2022physics,hu2022accelerating,choi2024accelerating}, data-based  model reduction \cite{song2016reduced,zhou2019reduced,tyrylgin2021multiscale}, and more recently, the data-free tensor decomposition technique \cite{Lu2025}. Another important aspect to consider is to reduce the cost for time integration. In spite of the various time integration schemes for related problems \cite{Shen2010, Wise2009, Hu2009,yang2017,zhao2017,shen2015movingcontact,gomez2023phase}, explicit type schemes are still popular for actual phase field simulations of AM microstructure \cite{Chadwick2022,xue2022physics}, due to their simplicity and the ease of implementation. However, it is known that explicit schemes could require very small time steps due to the stability concerns, leading to a large number of time increments when a realist time scale is considered for AM simulations. To the best of our knowledge, very little work has been reported on the use of alternative implicit or semi-implicit schemes for phase field simulations in AM, partially due to the complexity of the models and the potential physical constraints imposed by the rapid cooling rates in AM.   

In this work, we studied the feasibility of alternative time integration algorithms, especially a class of stabilized semi-implicit schemes, originally developed for the Allen-Cahn equation \cite{Allen1979} with constant coefficients and a relatively simple energy definition \cite{Shen2010}. For actual phase field simulations of AM microstructure evolution,  several difficulties have to be considered carefully.  For example, the phase field variable becomes multi-phase fields, a specific formulation needs to be derived for each phase field variable with appropriate stability conditions. The free energy defined for AM phase field models (e.g., \cite{Chadwick2021}) usually involves multi-source terms that couple several driving forces of different nature, which is non-convex  and may cause numerical stability issues. The coefficient  of equation (e.g., mobility) is not constant in the solid-liquid interfaces. To address these challenges, we develop the specific stabilized semi-implicit formulation with an appropriate stabilization coefficient suitable for AM phase field models, under the finite element framework and a revisited discrete energy law.  We used a phase field model \cite{Chadwick2021} specially designed for rapid solidification in AM processes for 316L stainless steel, as an example to illustrate our development. This model considered the solidification in a regime where the solid-liquid interface is moving very fast and there is absolute interfacial stability such that composition variations can be negligible, which can indeed occur in binary alloys at AM solidication rates, as reported for Fe-Cr \cite{Pinomaa2020} and Ni-Nb \cite{karayagiz2020finite} alloys. Validation of this model against experiments has been conducted recently \cite{Chadwick2025}. Therefore, we consider the adopted phase field model is appropriate to represent the rapid solidification condition in AM, although some of the effects, e.g., composition variations, were not considered. An overview of the model assumptions will be provided in the next section. We remark that the proposed computational framework is general and can be easily adapted to other phase field models. 

To exam the performance of the framework, we performed both 2D and 3D phase field simulations of grain growth and used the Rosenthal’s solution \cite{Rosenthal1946} to mimic the temperature evolution in a laser powder bed fusion AM process. The model parameters, including physical and kinetic parameters, are considered to be representative for AM 316L stainless steel and were chosen from the literature. Both the accuracy and energy stability of the numerical schemes were carefully examined. It is found that the proposed method can enable two orders-of-magnitude larger time steps than a traditional explicit scheme, without violating the energy stability requirement. The simulation results are  accurate and can reproduce the key observations of the experiments \cite{Chadwick2025} regarding different scan speeds and kinetic anisotropy effects.

This paper is organized as follows. Section 2 presents the key assumptions  and theory of the phase field model and the proposed computational framework. A detailed finite element formulation and the time integration schemes will be presented, along with a required discrete energy law and the energy stability analysis. Section 3 presents the 2D and 3D phase field simulation results. Finally, the paper closes with some concluding remarks. 

\section{The proposed framework}
\subsection{Model assumptions}
 {The phase field model adopted in our work was originally developed by \cite{Chadwick2021} under the following considerations and assumptions. 1) The model considers a binary Fe–Cr alloy system (the basic system of stainless steel), using a relatively large effective equilibrium distribution coefficient for austenite (a primary crystal structure of face-centered cubic type in stainless steel), and solute segregation is expected to be small. 2) The model considers solidification near the limit of absolute stability, where solute trapping becomes significant. Under these conditions, the solidification is expected to proceed predominantly with either planar interfaces or low-amplitude cells (negligible cellular substructures associated with solute segregation) \cite{Chadwick2021, Merchant1990}, which is consistent with theoretical calculations of interfacial velocity for 316L stainless steel under rapid solidification conditions \cite{Pinomaa2020}. 3) It is assumed that grains grow exclusively via epitaxial growth from pre-existing grains, with no new grain nucleation occurring during the process. 4) The solid-liquid interfacial energy is assumed to be isotropic, neglecting any orientation dependency of the interfacial energy. 5) The kinetic anisotropy is incorporated, by assuming that the solid-liquid interface velocity is strongly dependent on crystallographic orientation, i.e., interface mobility varies significantly with different crystal planes. This anisotropy plays a crucial role in competitive grain growth and morphological evolution during rapid solidification.}

\subsection{Phase field model of grain growth}
With the previous settings, the phase field model \cite{Chadwick2021} was developed for describing the rapid solidification in AM. Let us consider the following gradient flow problem
\begin{equation}
\label{eq:gradflow}
\frac{\partial\phi_i}{\partial t} =-M_i\frac{\delta E}{\delta \phi_i},
\end{equation}
where $\phi_i$ is the order parameter (OP)  {denoting grains of different crystal orientations}, i.e., a phase,  $M_i$ is the mobility of the $i$-th phase. In general, we can define the liquid phase by $i=0$, i.e., $\phi_0$, and the solid phase by $i\neq0$, i.e., $\phi_{i\neq0}$. We can see that the evolution of each OP (including liquid and solid phases) is driven by the free energy $E$. It usually includes a bulk potential and a gradient energy term and can be written as integral form over a domain $\Omega$
\begin{equation}
\label{eq:E}
E = \int_{\Omega}\mathcal{F}(\{\phi\}, \{\nabla\phi\})\, d\Omega = \int_{\Omega} \left[f\left(\{\phi\}, T\right) + \frac{\kappa}{2}\sum_{i=0}^N|\nabla \phi_i|^2 \right]\, d\Omega,
\end{equation}
where $\{\phi\}$ denotes the set of all $\phi_i$,  $\{\nabla\phi\}$ is the set of all $\nabla\phi_i$, $f\left(\{\phi\}, T\right)$ is the homogeneous free energy, and $\kappa$ is the gradient energy coefficient. The functional derivative of $E$, i.e., $\frac{\delta E}{\delta \phi_i}$, is defined according to the following definition
\begin{align}
\int_{\Omega} \frac{\delta E}{\delta \phi_i} \eta \, d\Omega
&= \int_{\Omega}\lim_{\epsilon \to 0}\frac{\mathcal{F}(\phi_i+\epsilon\eta, \nabla\phi_i + \epsilon\nabla\eta) - \mathcal{F}(\phi_i, \nabla\phi_i)}{\epsilon} \, d\Omega,
\end{align}
where \(\eta\) is assumed to be a sufficiently smooth function with a compact support in \(\Omega\), or more specifically, \(\eta \in H^1_0(\Omega)\) in our case. With a small perturbation $\epsilon \to 0$, we have
\begin{align}
\int_{\Omega} \frac{\delta E}{\delta \phi_i} \eta \, d\Omega \notag
&= \int_{\Omega} \frac{\partial \mathcal{F}}{\partial \phi_i} \eta + \frac{\partial \mathcal{F}}{\partial \nabla \phi_i} \nabla\eta \, d\Omega \notag \\
&= \int_{\Omega} \frac{\partial f}{\partial \phi_i} \eta + \kappa(\nabla \phi_i \cdot \nabla \eta) \, d\Omega \notag \\
&= \int_{\Omega}(\frac{\partial f}{\partial \phi_i}-\kappa\Delta\phi_i)\eta \, d\Omega,
\end{align}
where the divergence theorem is applied with the natural boundary condition ($\nabla \phi_i \cdot \boldsymbol{n}=0$ on $\partial \Omega$). We therefore obtain the Allen–Cahn type equation 
\begin{equation}
\label{eq:govern}
\frac{\partial\phi_i}{\partial t} + M_i\left(\frac{\partial f\left(\{\phi\}, T\right)}{\partial \phi_i} - \kappa\Delta\phi_i \right)=0.
\end{equation}
This provides a specific form of the gradient flow problem \eqref{eq:gradflow}. Now, we specify the explicit form of $f\left(\{\phi\}, T\right)$ with
\begin{equation}
\label{eq:free}
\begin{split}
f\left(\{\phi\}, T\right) =\ & W \left[ \sum_{i=0}^N \left(\frac{\phi_i^4}{4} - \frac{\phi_i^2}{2}\right) + \gamma_{PF} \sum_{i=0}^N \sum_{j>i}^N \phi_i^2 \phi_j^2 + \frac{1}{4} \right]  + L \frac{T_{Liq} - T}{T_{Liq}} h(\{\phi\}),
\end{split}
\end{equation}
where $W$ is the height of the multiwell energy, $\gamma_{PF}$ is the penalty coefficient that prevents OPs from overlap, $L$ is the latent heat, and $T_{Liq}$ is the  liquidus temperature. The last term of \eqref{eq:free} excluding the interpolation function $h(\{\phi\})$, i.e., $L \frac{T_{Liq} - T}{T_{Liq}} $, follows the Steinbach type formulation \cite{Steinbach2009}, which effectively shifts the energy minimum upward or downward depending on the deviation of the interface temperature from the melting point. Any deviation from the melting temperature energetically favors either the solid or the liquid phase. The interpolation function $h(\{\phi\})$ is defined as 
\begin{equation}
\label{eq:interph}
h(\{\phi\}) = \frac{p_0(\phi_0)}{ \sum_{i=0}^N p_i(\phi_i)},
\end{equation}
where $p_i(\phi_i)$ is the interpolation function for the $i$-th OP, and is a smooth interpolation function used to model the transition between the internal energy densities of the solid and liquid phases \cite{Wang1993}. Eq. \eqref{eq:interph} is a Moelans-type interpolation \cite{Moelans2011}, equal to zero in the solid and one in the liquid. It is normalized and smooth, providing thermodynamic consistency and improved numerical stability. A thermodynamically stable form for the interpolation function, according to \cite{Wang1993}, can be defined as
\begin{equation}
p_i(\phi_i) = \phi_i^3(20-45\phi_i + 36\phi_i^2 - 10\phi_i^3).
\end{equation}

These define the theoretical formulation of the phase field model adopted in our work. We remark that in the original work of \cite{Chadwick2021}, a transformation of variable was used to convert the OP into a pseudo signed distance function (PSDF), and a transformed formulation based on PSDF was used, instead of Eq. \eqref{eq:govern}. The model parameters were therefore defined based on the PSDF as well. In contrast, our work directly solves for the original OP using Eq. \eqref{eq:govern} with the model parameters defined in the following.

\subsection{Model parameters}
We define here the model parameters involved in the theoretical formulation of the phase field model. First, we define the gradient energy coefficient $\kappa$ as \cite{Moelans2008}
\begin{equation}
    \kappa = \frac{9\gamma_{SL}\zeta}{2}
\end{equation}
where $\zeta$ is the characteristic length, reflecting the thickness of the diffuse interface in the phase field model,  the constant $\gamma_{PF} = \frac{3}{2}$ and is related to the multiwell energy with $W=\frac{6\gamma_{SL}}{\zeta}$.

For the mobility $M_i$, we have a phase-dependent function with
\begin{equation}
\label{eq:mobility0}
M_{0} = L_{AC}(\{\phi\}),
\end{equation}
where $L_{AC}(\{\phi\})$ is defined as the weighted average over each pairwise interface mobility \cite{Moelans2008} and will be specified later. For the solid phase, 
\begin{equation}
M_{i \neq 0} =
\begin{cases}
    L_{AC}(\{\phi\}), & \phi_0 \geq \phi_c \\
    L_b^s, & \phi_0 < \phi_c,
\end{cases}
\end{equation}
where $L_b^s$ is a small bulk mobility, and $\phi_c$ is the lower bound of the liquid-solid interface. In practice, $L_{AC}$ is applied in the solid–liquid interface region to drive the evolution of the OP, while $L_b^s$ is used in the bulk solid region. A relatively large value of $L_{AC}$ is needed to capture the rapid dynamics at the solid–liquid interface. Following the work of \cite{Moelans2008}, we can define $L_{AC}(\{\phi\})$  as a weighted sum of individual pairwise interface mobilities
\begin{equation}
\label{eq:lac}
L_{AC}(\{\phi\}) = \frac{\sum_{i=0}^N \sum_{j>i}^N L_{ij}\phi_i^2 \phi_j^2}{\sum_{i=0}^N \sum_{j>i}^N \phi_i^2 \phi_j^2},
\end{equation}
where $L_{ij}$ is the phase field mobility between $\phi_i$ and $\phi_j$. This mobility formulation allows the model to resolve interface kinetics between each pair of phases through $L_{ij}$ while the weight $\phi_i^2\phi_j^2$ ensures that each mobility term is only active in the region where both phases coexist.  This structure accurately reflects the pairwise competition between phases. Furthermore, the normalization ensures consistency and boundedness of the effective mobility. The form of $L_{AC}$ indicates that it is nonzero only in the interface regions and zero within the bulk of each phase. For the liquid-solid pairs of OPs, the $L_{ij}$ is defined by
\begin{equation}
\label{eq:Lij}
L_{ij}\frac{3\zeta L}{T_{Liq}} =
\begin{cases}
    2\mu_{ij}, & \frac{\partial T}{\partial t} \leq 0 \\
    4\mu_0, & \frac{\partial T}{\partial t} > 0
\end{cases}
\end{equation}
where $\mu_{ij}$ is an orientation-dependent mobility coefficient, and $\mu_0$ represents the baseline magnitude of mobility.  As indicated by Eq. \eqref{eq:Lij}, a thicker interface ($\zeta$) and a larger latent heat ($L$) will result in a lower interface migration rate under a given thermal driving force, which aligns with the physical intuition. For solid–solid phase pairs, $L_{ij}$ is defined by
\begin{equation}
L_{ij} = min(L_{0k}, L_{k0}), \quad \text{for}\quad i\neq 0,\ j\neq 0.
\end{equation}
This equation assumes that the grain boundary (GB) mobility near the solid–liquid interface is comparable in magnitude to the solid–liquid interfacial mobility, allowing trijunctions to rotate. However, the GB mobility rapidly decreases in the bulk region, away from the melt pool interface. Consistent with \cite{Chadwick2021}, $L_b^s=\frac{\mu_0 T_{Liq}}{150\zeta L}$ is used to reduce grain coarsening while avoiding numerical artifacts. 

Finally, the orientation-dependent mobility coefficient $\mu_{ij}$ is defined as
\begin{equation}
\label{eq:anisoparam}
\mu_{ij}{(\boldsymbol{n}_{ij}}) = \mu_0 [1+\epsilon_4 (4(n_{1,ij}^4 + n_{2,ij}^4 + n_{3,ij}^4)-3)],
\end{equation}
where $\epsilon_4$ is the anisotropy parameter and ${\boldsymbol{n}_{ij}}= n_{1,ij}\boldsymbol{\hat{e_1}} + n_{2,ij}\boldsymbol{\hat{e_2}} + n_{3,ij}\boldsymbol{\hat{e_3}}$ denotes the normal vector of the solid-liquid interface in the coordinate frame of the solid domain,  {and may be rotated, using unit quaternions \cite{rowenhorst2015consistent}, with respect to the frame in which the preferred crystallographic  growth direction is defined}.  For each pair of OPs, the normal vector is computed as
\begin{equation}
{\boldsymbol{n}_{ij}} = \frac{\nabla \phi_i - \nabla \phi_j}{|\nabla \phi_i - \nabla \phi_j|}.
\end{equation}
In this way, the kinetic anisotropy of interface is considered in the present phase field model. In general,  {grains with preferred crystallographic orientation most aligned with the heat flow direction will be promoted with larger mobility coefficients $\mu_{ij}$, and a larger $\epsilon_4$ will lead to stronger grain selection and anisotropic growth}.

\subsection{Semi-discretized finite element formulation}
In our work, we use the finite element (FE) method to solve the phase field equation \eqref{eq:govern}. We first derive the weak form of the problem. Given that the natural boundary condition applies to the problem, we have
\begin{equation}
\label{eq:weak}
\int_{\Omega}\delta \phi \frac{\partial \phi_i}{\partial t} \, d\Omega + \int_{\Omega}\delta \phi M_i \frac{ \partial f\left(\{\phi\}, T\right)}{\partial \phi_i} \, d\Omega + \int_{\Omega}\nabla \delta \phi \cdot M_i \kappa \nabla\phi_i \, d\Omega=0, \quad \forall t \in [0, t_f],
\end{equation}
where  $\delta \phi$ is a test function, $t_f$ is the final time. Eq.\eqref{eq:weak} can be discretized using a FE approximation of OP
\begin{equation}
\phi_h = \sum_{p\in S}N_p\phi_p=\boldsymbol{N}\boldsymbol{\phi},
\end{equation}
where $N_p$ is the FE shape function,  $S$ represents the set of supporting nodes of the FE shape functions, $\phi_p$ is the nodal solution, $\boldsymbol{N}$ and $\boldsymbol{\phi}$ are the nodal vectors of the $N_p$ and $\phi_p$. Considering Eq. \eqref{eq:weak} holds for the arbitrary $\delta \phi$ and $\delta \phi_h=\boldsymbol{N}\delta\boldsymbol{\phi}$, we then have 
\begin{equation}
\label{eq:FEsemi}
\int_{\Omega}\boldsymbol{N}^T \frac{\partial \phi_i}{\partial t} \, d\Omega + \boldsymbol{Q} ({\{\boldsymbol{\phi}\}, \boldsymbol{\phi}_i}) + \boldsymbol{K}(\{\boldsymbol{\phi}\})\ \boldsymbol{\phi}_i =0, \quad \forall t \in [0, t_f],
\end{equation}
with 
\begin{align}
     & \boldsymbol{Q} ({\{\boldsymbol{\phi}\}}, \boldsymbol{\phi}_i)=\int_{\Omega}\boldsymbol{N}^T M_i(\{\boldsymbol{\phi}\}) \frac{ \partial f\left(\{\boldsymbol{\phi}\}, T\right)}{\partial {\phi}_i} \, d\Omega,\\
     &\boldsymbol{K}(\{\boldsymbol{\phi}\}) =  \int_{\Omega}\boldsymbol{B}^T M_i(\{\boldsymbol{\phi}\}) \ \kappa \boldsymbol{B}\, d\Omega,
\end{align}
where $\boldsymbol{\phi}_i$ is the nodal vector of $i$-th OP, $\{\boldsymbol{\phi}\}$ is the set of all nodal vectors, $\boldsymbol{B}$ is the derivative of the shape function. This is the semi-discrete form of the problem and is a nonlinear equation due to the definitions of $f$ and $M_i$. To solve this problem, an appropriate time integration scheme is needed.

Conventional time integration schemes include explicit, implicit, semi-implicit ones. However, fully explicit and implicit methods both have some disadvantages in terms of time step constraints or convergence concerns due to the nonlinearity. The semi-implicit formulation seems a good choice, but should be carefully designed in order to account for the discrete energy law described in the next. 

\subsection{Discrete energy law}
Given the nature of gradient flow problems, the solutions of phase field models are expected to satisfy an energy law. This should be guaranteed by a time integration method. Now, taking $L^2$ inner product of Eq. \eqref{eq:govern} with $(\frac{\partial f}{\partial \phi_i} - \kappa\Delta\phi_i)$, we can find that
\begin{equation}
\int_{\Omega}\frac{\partial \phi_i}{\partial t}(\frac{\partial f}{\partial \phi_i} - \kappa\Delta\phi_i) \, d\Omega  = - \int_{\Omega} M_i\left(\frac{\partial f}{\partial \phi_i} - \kappa\Delta\phi_i \right)^2 \, d\Omega.
\end{equation}
On the other hand, taking the time derivative of $E$, i.e., Eq. \eqref{eq:E}, yields
\begin{align}
\label{eq:energy}
\frac{d E}{d t} \notag
&= \int_{\Omega} \left( \sum_{i=0}^N \frac{\partial f}{\partial \phi_i} \frac{\partial \phi_i}{\partial t} + \frac{\partial f}{\partial T} \frac{\partial T}{\partial t} + \sum_{i=0}^N\kappa \nabla \phi_i \cdot \nabla \frac{\partial \phi_i}{\partial t} \right) \, d\Omega \notag \\
&= \sum_{i=0}^N \left[  \int_{\Omega} \left( \frac{\partial f}{\partial \phi_i} \frac{\partial \phi_i}{\partial t} - \kappa \Delta\phi_i \frac{\partial \phi_i}{\partial t} \right) \,  d\Omega + \int_{\partial\Omega} \frac{\partial \phi_i}{\partial t} \nabla \phi_i \cdot \boldsymbol{n} \, d\partial\Omega \right] + \int_{\Omega}\frac{\partial f}{\partial T} \frac{\partial T}{\partial t} \, d\Omega \notag \\
&= \sum_{i=0}^N\int_{\Omega} \frac{\partial \phi_i}{\partial t} \left( \frac{\partial f}{\partial \phi_i} - \kappa \Delta \phi_i \right) \, d\Omega + \int_{\Omega}\frac{\partial f}{\partial T} \frac{\partial T}{\partial t} \, d\Omega \notag \\
&= - \sum_{i=0}^N\int_{\Omega} M_i\left(\frac{\partial f}{\partial \phi_i} - \kappa\Delta\phi_i \right)^2 \, d\Omega + \int_{\Omega}\frac{\partial f}{\partial T} \frac{\partial T}{\partial t} \, d\Omega .
\end{align}
This indicates that the total free energy of the system evolves with both the OPs and the temperature evolution. The energy does not necessarily decrease over time, if temperature increases, as reflected by the second term in the above equation. This is different from conventional discrete energy law for the Allen-Cahn type equation.

To establish a suitable energy law for the phase field model, we can focus on the first term in the last line of the Eq. \eqref{eq:energy}, which reads 
\begin{equation}
 - \sum_{i=0}^N \int_{\Omega} M_i\left(\frac{\partial f}{\partial \phi_i} - \kappa\Delta\phi_i \right)^2 \, d\Omega = \left.\frac{dE}{dt}\right|_{\frac{\partial f}{\partial T} \frac{\partial T}{\partial t} =0} \leq 0.    
\end{equation}
This implies that 
\begin{equation}
 \left.\frac{dE}{dt}\right|_{\frac{\partial f}{\partial T} =0} \leq 0,  
\end{equation}
and 
\begin{equation}
 \left.\frac{dE}{dt}\right|_{\frac{\partial T}{\partial t}=0} \leq 0.  
\end{equation}
The last equation implies that if the energy $E$ is computed with the same temperature, we should have the following discrete energy law
\begin{equation}
\label{eq:energylaw}
E(\{\phi^{k+1}\},T^{k+1}) - E(\{\phi^{k}\},T^{k+1})\leq 0,
\end{equation}
where $k$ refers the $k$-th time step. This revisited energy law allows us to verify the energy stability of the solution and is generally applicable to various phase field models that involve temperature driving forces. This revisited discrete energy law will be  used as the energy stability condition in the numerical examples. And the time integration algorithms presented in the next are expected to guarantee this discrete energy stability.

\subsection{Stabilized semi-implicit time integration}
As mentioned earlier and in the introduction, we are interested in a class of semi-implicit time integration schemes that could allow the use of larger time steps for phase field simulations, and more specifically, for solving the semi-discrete phase field model \eqref{eq:FEsemi} or similar types. Additionally, the discrete energy law should be naturally and unconditionally enforced. 

Let us first consider a conventional semi-implicit scheme
\begin{equation}
\int_{\Omega}\boldsymbol{N}^T \boldsymbol{N} \frac{\boldsymbol{\phi}_i^{k+1}-\boldsymbol{\phi}_i^{k}}{\Delta t} \, d\Omega\ +  \boldsymbol{Q} ({\{\boldsymbol{\phi}^k\}, \boldsymbol{\phi}^k_i}) + \boldsymbol{K}(\{\boldsymbol{\phi}^k\})\ \boldsymbol{\phi}^{k+1}_i 
=0, 
\end{equation}
where $\boldsymbol{\phi}_i^{k+1}$ and $\boldsymbol{\phi}_i^k$ represent the solutions at the current and previous time steps, respectively, and $\Delta t = t^{k+1}-t^k$. This semi-implicit time integration scheme could be used for phase field simulations. However, it does not grantee the energy stability for arbitrarily large time steps, as stated in \cite{Shen2010,Lu2025}.  To overcome the time step constraint, we can adopt a stabilized semi-implicit scheme, which reads
\begin{equation}
\int_{\Omega}\boldsymbol{N}^T \boldsymbol{N} (\boldsymbol{\phi}_i^{k+1}-\boldsymbol{\phi}_i^{k})(\frac{1}{\Delta t} + \alpha M_i (\{\boldsymbol{\phi}^k\})) \, d\Omega + \boldsymbol{Q} ({\{\boldsymbol{\phi}^k\}, \boldsymbol{\phi}^k_i}) + \boldsymbol{K}(\{\boldsymbol{\phi}^k\})\ \boldsymbol{\phi}^{k+1}_i =0, 
\end{equation}
where $\alpha$ is a stabilizing coefficient.  An appropriate $\alpha$ needs to be obtained to balance the stabilization effect and the accuracy. One way to calculate this  $\alpha$ will be presented in the next section, along with the energy stability analysis. As a result, it is expected that the stabilized semi-implicit formulation can enable arbitrarily large time steps without violating the revisited energy law \eqref{eq:energylaw}. The final discrete equation for the phase field model is then
\begin{equation}
\label{eq:firstorder}
\boldsymbol{M} \boldsymbol{\phi}_i^{k+1}- \boldsymbol{M}\boldsymbol{\phi}_i^{k} 
+  \boldsymbol{Q} ({\{\boldsymbol{\phi}^k\}, \boldsymbol{\phi}^k_i}) + \boldsymbol{K}(\{\boldsymbol{\phi}^k\})\ \boldsymbol{\phi}^{k+1}_i =0, 
\end{equation}
with 
\begin{equation}
\boldsymbol{M} =\int_{\Omega}\boldsymbol{N}^T \boldsymbol{N} (\frac{1}{\Delta t} + \alpha M_i (\{\boldsymbol{\phi}^k\})) \, d\Omega 
\end{equation}
This is a first-order stabilized semi-implicit scheme. Following the derivation of \cite{Shen2010}, we can derive a second-order stabilized scheme as
\begin{align}
& \int_{\Omega}\boldsymbol{N}^T \boldsymbol{N} (3\boldsymbol{\phi}_i^{k+1}-4\boldsymbol{\phi}_i^{k} + \boldsymbol{\phi}_i^{k-1})(\frac{1}{2\Delta t} ) \, d\Omega \notag \\
& + \int_{\Omega}\boldsymbol{N}^T \boldsymbol{N} (\boldsymbol{\phi}_i^{k+1}-2\boldsymbol{\phi}_i^{k}+\boldsymbol{\phi}_i^{k-1}) \alpha M_i (\{\boldsymbol{\phi}^k\})  \, d\Omega \notag \\
& +2\boldsymbol{Q} ({\{\boldsymbol{\phi}^k\}, \boldsymbol{\phi}^k_i})- \boldsymbol{Q} ({\{\boldsymbol{\phi}^{{k-1}}\}, \boldsymbol{\phi}^{k-1}_i}) \notag\\
& + \boldsymbol{K}(\{\boldsymbol{\phi}^k\})\ \boldsymbol{\phi}^{k+1}_i=0,
\end{align}
and the discrete form with the second-order time integration is 
\begin{align}
\label{eq:secondorder}
& (\frac{3}{2}\boldsymbol{M}_t+\boldsymbol{M}_\alpha) \boldsymbol{\phi}_i^{k+1}- 2\boldsymbol{M}\boldsymbol{\phi}_i^{k} + (\frac{1}{2}\boldsymbol{M}_t+\boldsymbol{M}_\alpha)\boldsymbol{\phi}_i^{k-1} 
+2\boldsymbol{Q} ({\{\boldsymbol{\phi}^k\}, \boldsymbol{\phi}^k_i}) - \boldsymbol{Q} ({\{\boldsymbol{\phi}^{{k-1}}\}, \boldsymbol{\phi}^{k-1}_i}) + \boldsymbol{K}(\{\boldsymbol{\phi}^k\})\ \boldsymbol{\phi}^{k+1}_i =0, 
\end{align}
with
\begin{align}
&\boldsymbol{M}_t =\int_{\Omega}\boldsymbol{N}^T \boldsymbol{N} \frac{1}{\Delta t}  d\Omega, \\
&\boldsymbol{M}_\alpha =\int_{\Omega}\boldsymbol{N}^T \boldsymbol{N}  \alpha M_i (\{\boldsymbol{\phi}^k\}) \, d\Omega. 
\end{align}

Both the first-order \eqref{eq:firstorder} and second-order \eqref{eq:secondorder} schemes are expected to be energetically stable in the sense of Eq. \eqref{eq:energylaw}. However, it is important to note that being energetically stable does not imply that the time step $\Delta t$ can be arbitrarily large without affecting the accuracy of the solution. In practice, a sufficiently small time step is  required to accurately capture the temperature history in AM processes.  We can expect that this time step is still relatively very large compared to the one required by an explicit scheme, as shown later in the numerical examples. Compared to the fully implicit schemes, the semi-implicit formulation does not rely on an iterative procedure and can avoid convergence issues for solving the underlying nonlinear (and nonconvex) phase field equation, leading to an efficient and easy-to-implement framework. 

\subsection{Energy stability analysis}
To ensure the energy stability in the proposed semi-implicit schemes, the stabilizing coefficient $\alpha$ can be chosen as 
\begin{equation}
\alpha \geq \frac{1}{2}\sup(\frac{ \partial^2 f\left(\{\phi\}, T\right)}{\partial \phi_i^2}), \quad \forall \phi_i \in [0,1]
\end{equation}
The is obtained by following the proofs in \cite{Shen2010,Lu2025}. However, this is not a good choice for the phase field models in AM. First, the proofs assumed the mobility $M_i$ is a constant over the solution domain, which might  be valid only for the solid-solid interfaces in our phase field models. Second, this choice would lead to an extremely large $\alpha$ that can  significantly affect the accuracy of the solution.  Therefore, we propose the following equation to calculate a suitable stabilizing coefficient
\begin{equation}
\label{eq:modifysup}
\alpha \geq \frac{1}{2}\sup(\frac{ \partial^2 \tilde{f}\left(\{\phi\}, T\right)}{\partial \phi_i^2}), \quad \forall \phi_i \in [0,1],
\end{equation}
where   $\tilde{f}$ is a modified free energy functional with respect to Eq. \eqref{eq:free}   {and can be derived from the following analysis. Let us consider the discrete energy law Eq. \eqref{eq:energylaw}, which can be written as}
 {
\begin{align}
\label{eq:energylaw_split}
& \underbrace {E(\{\phi^{k+1}\},T^{k+1})_{\Omega_{ss}} - E(\{\phi^{k}\},T^{k+1})_{\Omega_{ss}}}_{\text{solid phase}} +\underbrace { E(\{\phi^{k+1}\},T^{k+1})_{\Omega_{sl}} - E(\{\phi^{k}\},T^{k+1})_{\Omega_{sl}}}_{\text{solid-liquid interface}} \leq 0,
\end{align}
where $E(\cdot)_{\Omega_{ss}}$ denotes the energy over the solid phase, $E(\cdot)_{\Omega_{sl}}$ is that on the solid-liquid interface. We can first focus on the second part (solid-liquid interface) of the above equation and try to prove that
\begin{equation}
\label{eq:slinterface}
E(\{\phi^{k+1}\},T^{k+1})_{\Omega_{sl}} - E(\{\phi^{k}\},T^{k+1})_{\Omega_{sl}}\leq0.
\end{equation}
By analyzing the phase field equation \eqref{eq:govern} and the free energy definition \eqref{eq:free},  we can see that the governing equation on the solid-liquid interface can be written as  
\begin{align}
& \frac{\partial\phi_i}{\partial t} = - M_i\left(\frac{\partial f\left(\{\phi\}, T\right)}{\partial \phi_i} - \kappa\Delta\phi_i \right) \approx  -M_i\left(L\frac{T_{Liq}-T}{T_{Liq}}\frac{\partial h(\{\phi\})}{\partial \phi_i} \right), \quad \forall x\in \Omega_{sl}.
\end{align}
This is based on the fact that the temperature driving force is dominant on the solid-liquid interface. Therefore, given the positivity of $M_i, L,$, and $L\frac{T_{Liq}-T}{T_{Liq}}$ during solidification,   the evolution of the OP $\phi_i$ is related to  the sign of $\frac{\partial h(\{\phi\})}{\partial \phi_i}$. Now, we should have
\begin{equation}
\frac{\partial\phi_i}{\partial t} \geq 0, \quad i\neq 0, 
\end{equation}
during solidification, meaning the growth of grains, and we must have $\frac{\partial h(\{\phi\})}{\partial \phi_i}\leq0$, which is consistent with the definition of $h(\{\phi\})$. This forces the decrease of $h(\{\phi\})$ and consequently a monotonic decrease of the free energy on the solid-liquid interface, as $f(\cdot)_{\Omega_{sl}}\approx L \frac{T_{Liq} - T}{T_{Liq}} h(\{\phi\})$, which confirms Eq. \eqref{eq:slinterface}.} 

 {This analysis can be done for melting/re-melting conditions as well. For example, let us assume a melting condition, i.e., the decrease of OP:  $\frac{\partial\phi_i}{\partial t} < 0,\quad i\neq 0$. In this case, we should have $ h(\{\phi^{k+1}\}) - h(\{\phi^{k}\})>0$,  based on the definition of $h(\{\phi\})$.  Additionally, considering the temperature under melting conditions: $T>T_{Liq}$, we should have ${f(\phi^{k+1})_{\Omega_{sl}}}-{f(\phi^{k})_{\Omega_{sl}}}\approx L \frac{T_{Liq} - T}{T_{Liq}} (h(\{\phi^{k+1}\}) - h(\{\phi^{k}\}))<0$,  meaning the decrease of the free energy.  If a separate material melting point (different from $T_{Liq}$) is considered in the phase field model under melting conditions, we can still have a similar result. However, this would require a modification of the current phase field model and is out of the scope of this work.}

 {Therefore, we can simply impose 
\begin{equation}
\label{eq:ssinterface}
E(\{\phi^{k+1}\},T^{k+1})_{\Omega_{ss}} - E(\{\phi^{k}\},T^{k+1})_{\Omega_{ss}}\leq0,
\end{equation}
to ensure the global energy stability by choosing the stabilizing coefficient from Eq. \eqref{eq:modifysup} for solid-solid interfaces and considering that
\begin{equation}
    \tilde{f}=\ W \left[ \sum_{i=0}^N \left(\frac{\phi_i^4}{4} - \frac{\phi_i^2}{2}\right) + \gamma_{PF} \sum_{i=0}^N \sum_{j>i}^N \phi_i^2 \phi_j^2 + \frac{1}{4} \right]. 
\end{equation}
Given the above definition, we have
\begin{equation}
\frac{ \partial^2 \tilde{f}\left(\{\phi\}, T\right)}{\partial \phi_i^2} = W [3\phi_i^2 -1 + \gamma_{PF} \sum_{j\neq i}^N 2\phi_j^2] = W [\sum_{j=0}^N3\phi_j^2 -1],
\end{equation} 
where $\gamma_{PF}=3/2$ is applied for the second equality. Furthermore, we can expect that $\sum_{j=0}^N\phi_j^2\leq 1$, given the physical constraint of the problem. Hence, we can have a simple choice for the stabilizing coefficient from Eq. \eqref{eq:modifysup}, which reads
\begin{equation}
\alpha \geq W.
\end{equation}
This choice of $\alpha$ can apply to both the first-order \eqref{eq:firstorder} and second-order \eqref{eq:secondorder} schemes and is expected to guarantee the energy stability \eqref{eq:energylaw} and \eqref{eq:energylaw_split} over the entire computational domain. The numerical results in the next section will further confirm this choice of stabilizing coefficient and the theoretical energy stability analysis. }

\section{Numerical Experiments}
\subsection{Initial grain structure and implementation details}
In our work, the temperature evolution is assumed given and is defined by the Rosenthal’s solution \cite{Rosenthal1946} to mimic a laser powder bed fusion AM process. The detailed equation describing the temperature profile and the related parameters  are provided in  { \ref{sec:App1}}. The initial grain structure is generated using the Centroidal Voronoi Tessellation (CVT) method \cite{CVT1999}. To improve the uniformity of the CVT generated  grain structure, five iterations of Lloyd’s algorithm \cite{Lloyd1982} are performed. A grain orientation ID is then randomly assigned to each CVT generated grain.  The physical and kinetic parameters are considered to be representative for 316L stainless steel and are summarized in \tablename~\ref{tab:param}. These parameters are assumed to be similar to that of the austenitic $\gamma$-phase of pure iron \cite{Chadwick2021, Sun2004}.

\begin{table}[h]
\centering
\caption{Physical and kinetic parameters}
\begin{tabular}{lll}
\toprule
\textbf{Parameter} & \textbf{Value} & \textbf{Reference} \\
\midrule
$L$ & $1.9 \times 10^{-9}\ \text{J}/ \,\upmu\text{m}^3$  & \cite{Kim1975} \\
$T_{Liq}$ & $1700 \ \text{K}$ & \cite{Kim1975} \\
$\gamma_{SL}$ & $2 \times 10^{-13}\ \text{J}/ \,\upmu\text{m}^2$ & \cite{Sun2004} \\
$\mu_{0}$ & $2.17 \times 10^{5}\ \upmu\text{m}/ (\text{s}.\text{K}) $ & \cite{Sun2004} \\
$\epsilon_{4}$ & 0.11 & \cite{Sun2004} \\
\bottomrule
\end{tabular}
\label{tab:param}
\end{table}

The overall computational framework for phase field simulations is implemented using Python packages \cite{Python}. This includes the generation of the initial grain structure,  the finite element model, and time integration schemes.  The hardware used for this work is the Intel Xeon Gold 6548Y+ Processor with 32 CPU cores and 512 GB RAM.

\subsection{Energy stability}
\label{sec:energystability}
\begin{figure}[htbp]
  \centering
  \includegraphics[width=0.38\textwidth]{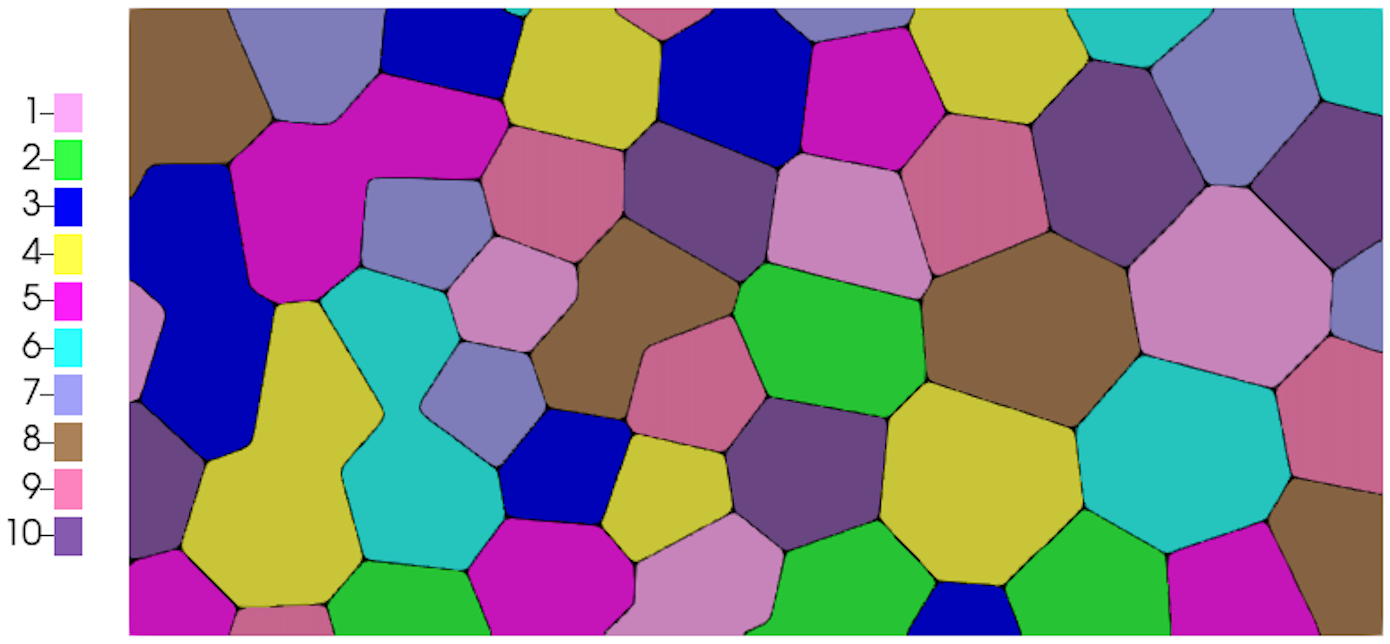}
  \caption{Initial grain structure with 10 randomly assigned orientation IDs}
  \label{Lbs0}
\end{figure}

We first verify whether the computational framework, particularly the stabilized semi-implicit scheme, can ensure the energy stability under various kinetic conditions.  {Meanwhile, we aim to demonstrate how energy stability influences the accuracy of the simulated grain structure, thereby highlighting the necessity of energy-stable schemes in grain growth simulations.} To this end, a 2D domain of size \(76.8\,\upmu\text{m} \times 38.4\,\upmu\text{m}\) is selected and the initial grain structure is generated with 10 random orientations,  {and we use 10 OPs to represent the different grains}, as illustrated in \figurename~\ref{Lbs0}. The mesh size is \(640 \times 320\). The characteristic length $\zeta = 0.204 \, \upmu \text{m}$, which is 1.7 times of the element size. Two different mobility values are selected in our tests: $M_i=6.34 \times10^{15} \ \upmu\text{m}^3/(\text{J}.\text{s})$ and $M_i=4.25 \times10^{17} \ \upmu\text{m}^3/(\text{J}.\text{s})$. The first one corresponds to a representative value of $L_b^s$, while the second one corresponds to a representative value of $L_{\mathrm{AC}}$ calculated by Eq. \eqref{eq:lac}. For each case, simulations are conducted  with pure solid phases under the isothermal condition with relatively large time steps for both the first- and second-order stabilized semi-implicit schemes with $\alpha=5.88 \times 10^{-12}$. They are compared with those without the stabilizing term ($\alpha=0$). 

\begin{figure}[htbp]
  \centering
  \subcaptionbox{\,Reference solution at 0.01\,s}[0.35\textwidth]{%
    \includegraphics[width=\linewidth]{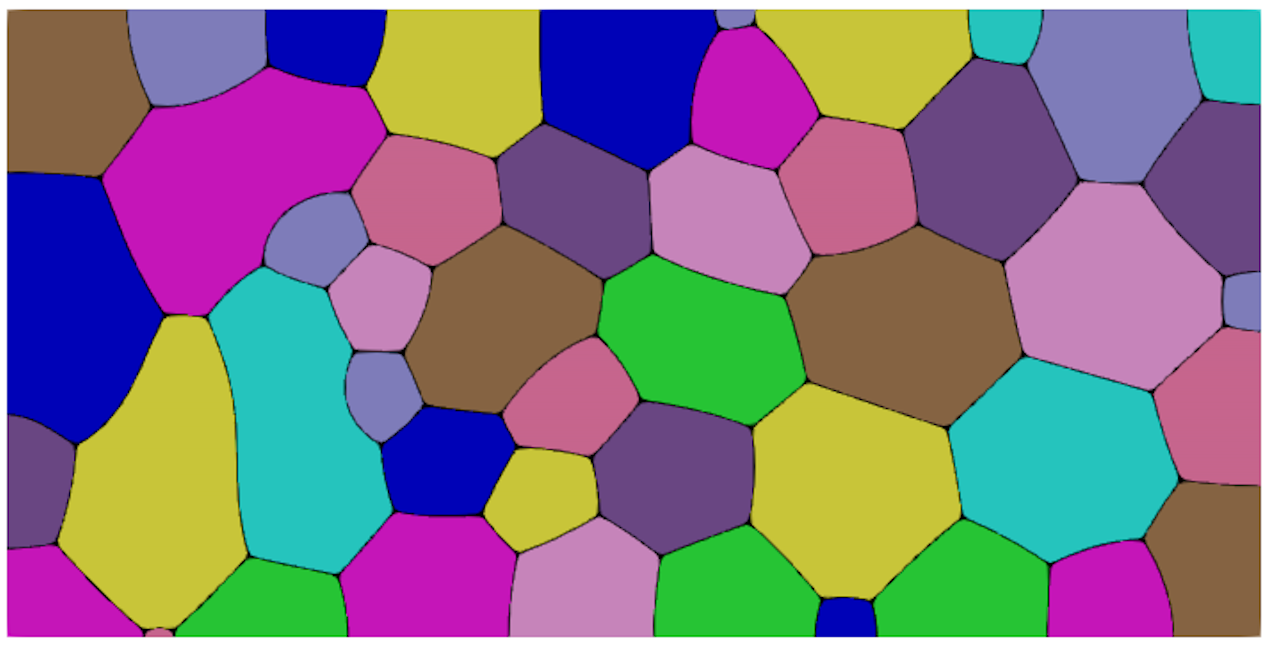}
  } \\
  \centering
   \vspace{5pt} 
  \subcaptionbox{\,1st order unstab. semi-impl.}[0.35\textwidth]{%
  \includegraphics[width=\linewidth]{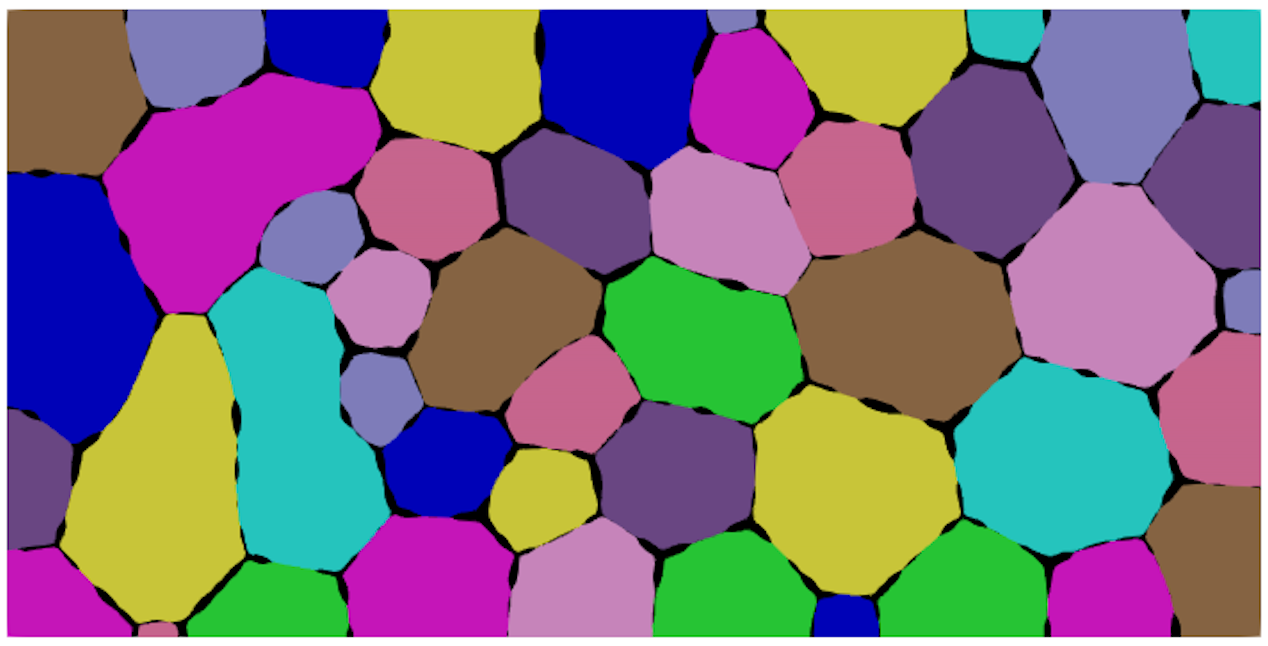}
  }
  \hspace{0.03\textwidth}
  \subcaptionbox{\,1st order stab. semi-impl.}[0.35\textwidth]{%
    \includegraphics[width=\linewidth]{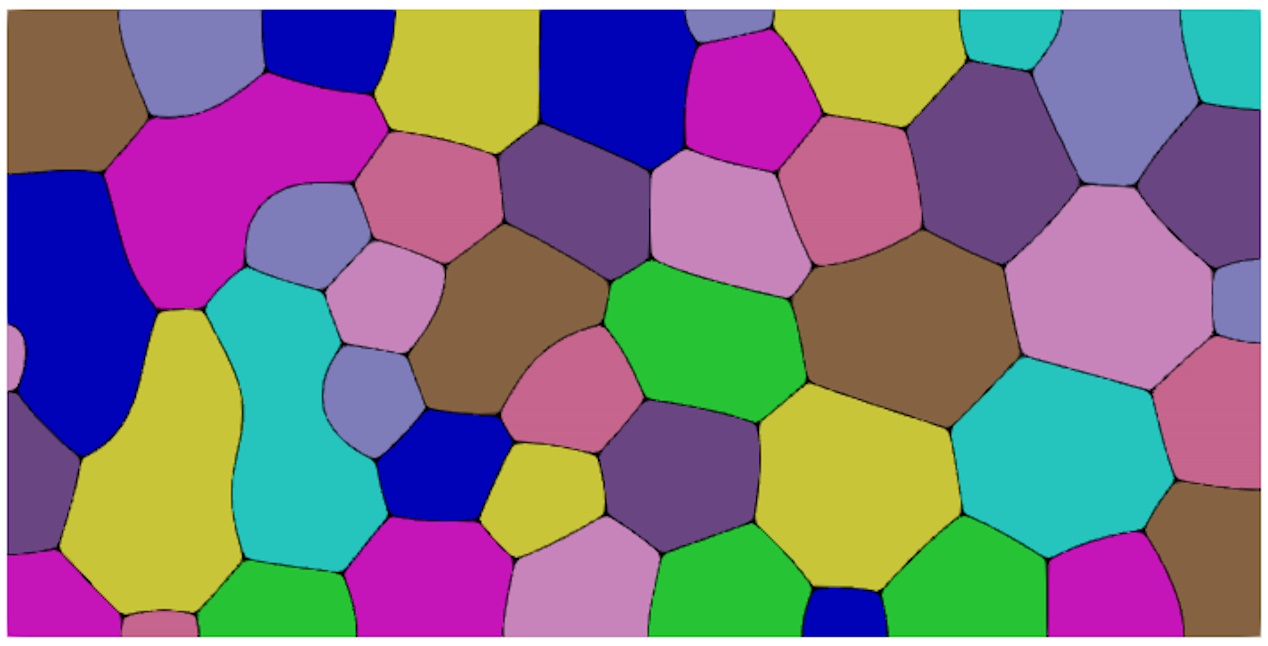}
  } \\
  \vspace{5pt} 
  \centering
  \subcaptionbox{\,2nd order unstab. semi-impl.}[0.35\textwidth]{%
    \includegraphics[width=\linewidth]{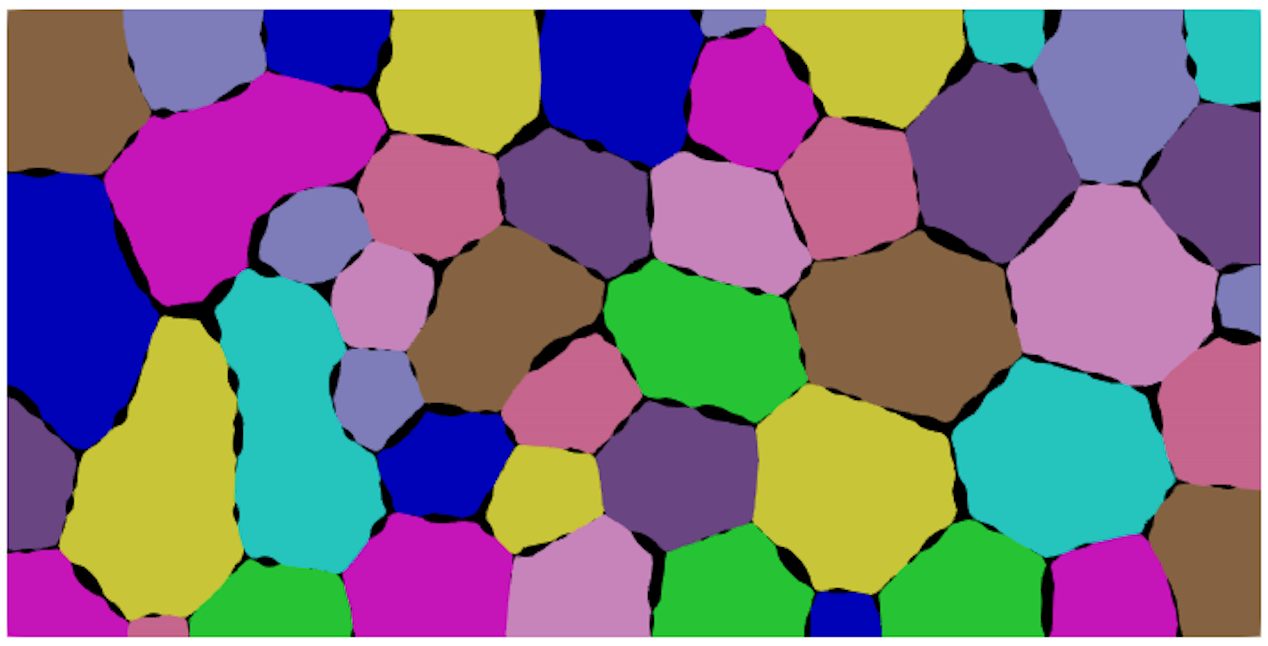}
  }
  \hspace{0.03\textwidth}
  \subcaptionbox{\,2nd order stab. semi-impl.}[0.35\textwidth]{%
    \includegraphics[width=\linewidth]{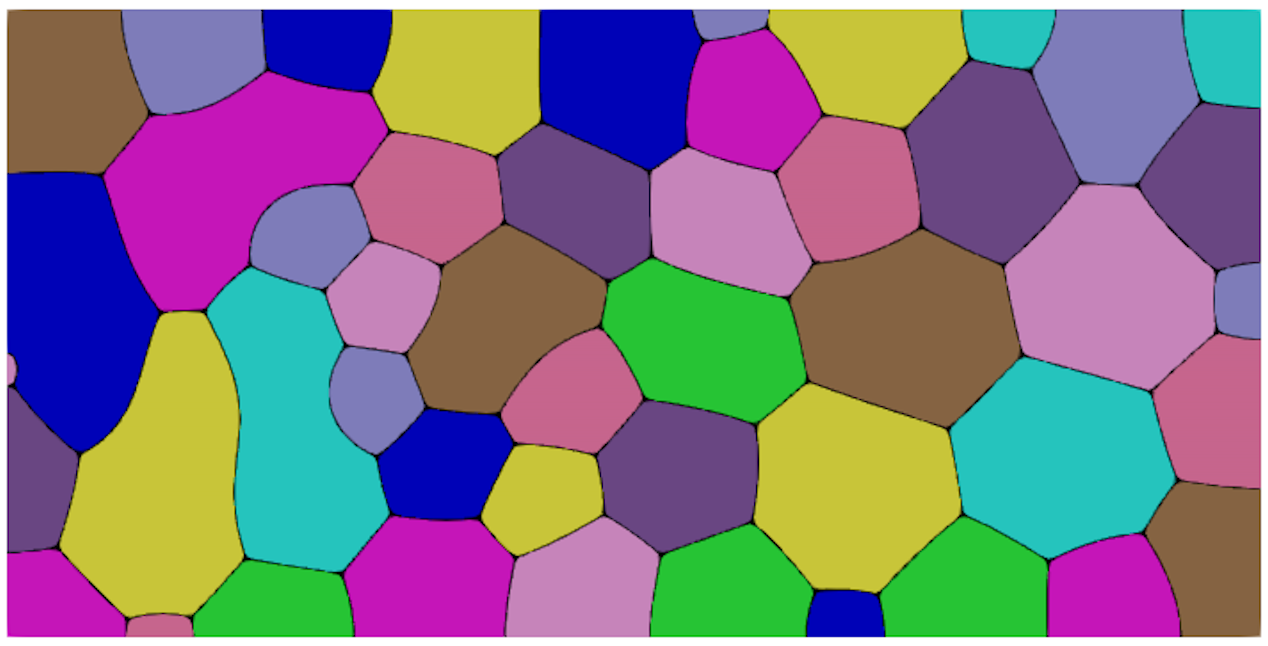}
  }  
  \caption{Grain structure for $M_i  = 6.34 \times10^{15} \ \upmu\text{m}^3/(\text{J}.\text{s})$. (b)  (c): \(\Delta t = 5 \times 10^{-5}\) s, (d)  (e): \(\Delta t = 1 \times 10^{-4}\) s.}
  \label{Lbs1}
\end{figure}

Figure~\ref{Lbs1}(a) shows the reference phase field solution at 0.01\,s obtained with a  {relatively} small time step  {($\Delta t = 5 \times 10^{-7}$ s)} and the mobility $M_i =  6.34 \times10^{15} \ \upmu\text{m}^3/(\text{J}.\text{s})$. Figure~\ref{Lbs1}(b)(c) show the results obtained with the first-order semi-implicit schemes with a time step size $\Delta t = 5 \times 10^{-5}$ s. It is clear that the unstabilized semi-implicit scheme would lead to  a strong instability on the GBs, whereas the stabilized semi-implicit scheme remains robust on capturing the movement of GBs with the large time step. Similar results can be drawn from the results of second-order semi-implicit schemes, as in Figure~\ref{Lbs1}(d)(e). In addition, we can notice that the second-order scheme could produce slightly more accurate results, compared to the first-order schemes.  {For example, we can focus on the blue grain next to the left boundary, and compare the results in \figurename~\ref{Lbs1}(a)(c)(e). The result from the second-order scheme looks more similar to the reference one, whereas the first-order scheme slightly delayed the evolution of the solution. This difference may be reduced by adjusting the time steps.}

\begin{figure}[htbp]
  \centering
  \subcaptionbox{\,Reference solution at \(1.6 \times 10^{-4}\) s}[0.35\textwidth]{%
    \includegraphics[width=\linewidth]{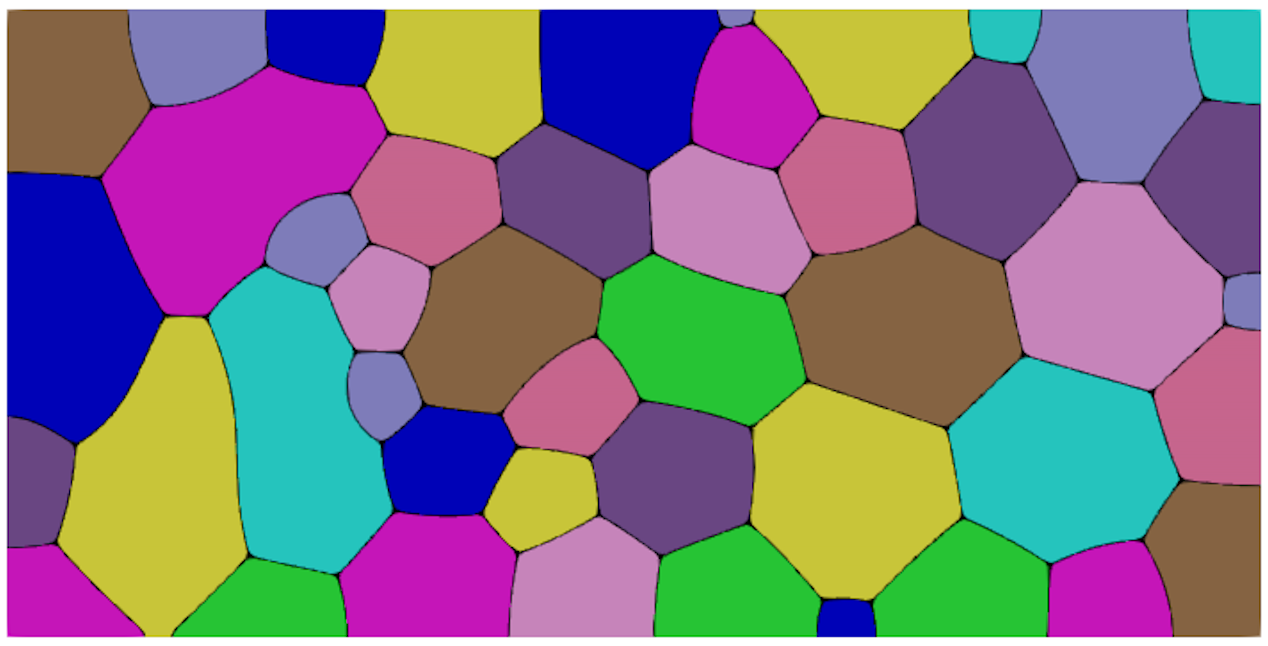}
  }\\
  \centering
   \vspace{5pt} 
  \subcaptionbox{\,1st order unstab. semi-impl.}[0.35\textwidth]{%
    \includegraphics[width=\linewidth]{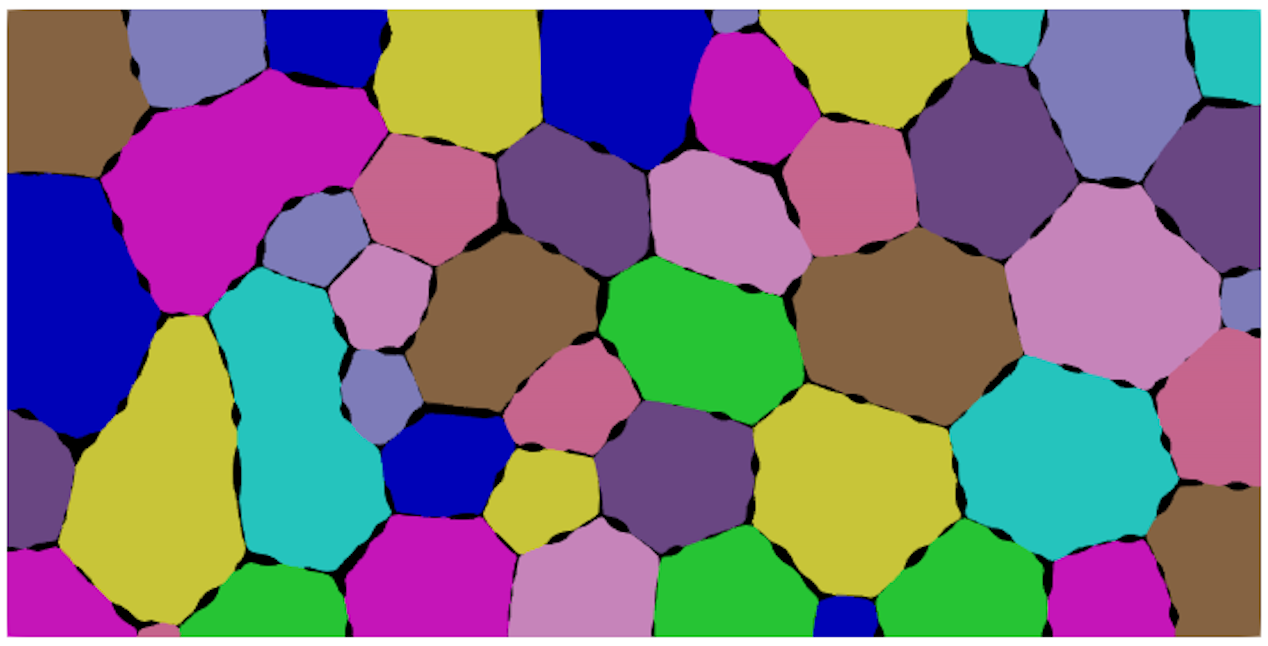}
  }
  \hspace{0.03\textwidth}
  \subcaptionbox{\,1st order stab. semi-impl.}[0.35\textwidth]{%
    \includegraphics[width=\linewidth]{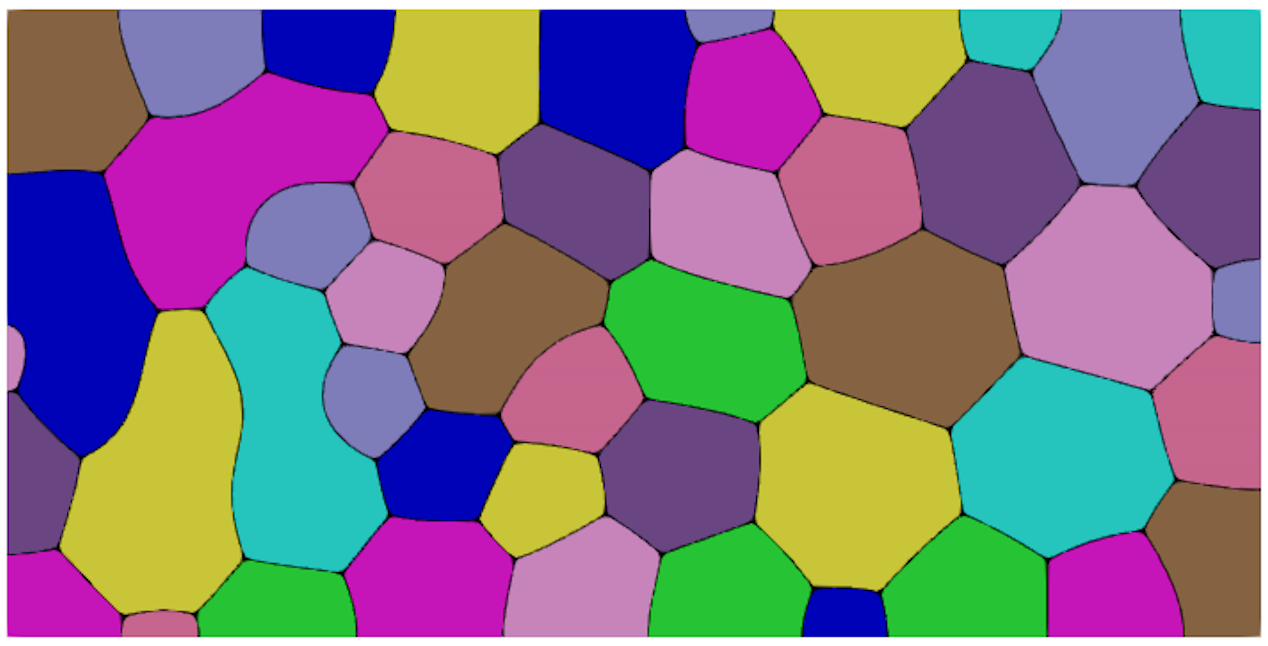}
  }\\
 \centering
   \vspace{5pt} 
  \subcaptionbox{\,2nd order unstab. semi-impl.}[0.35\textwidth]{%
    \includegraphics[width=\linewidth]{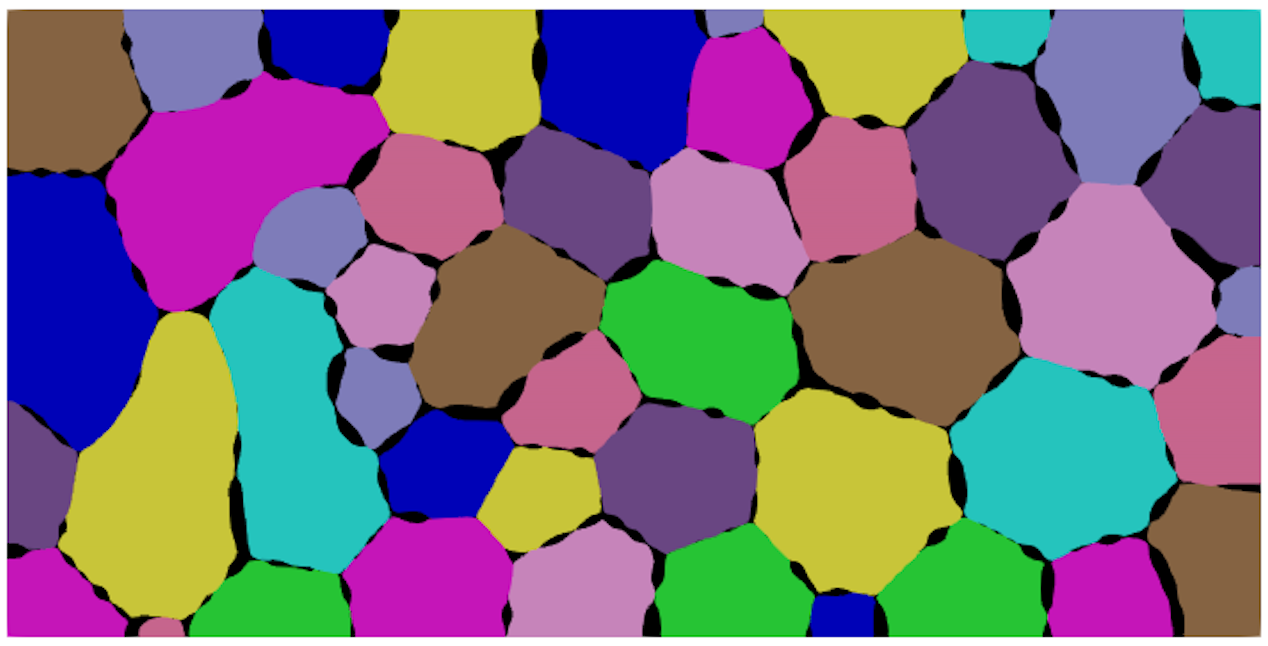}
  }  
  \hspace{0.03\textwidth}
  \subcaptionbox{\,\,2nd order stab. semi-impl.}[0.35\textwidth]{%
    \includegraphics[width=\linewidth]{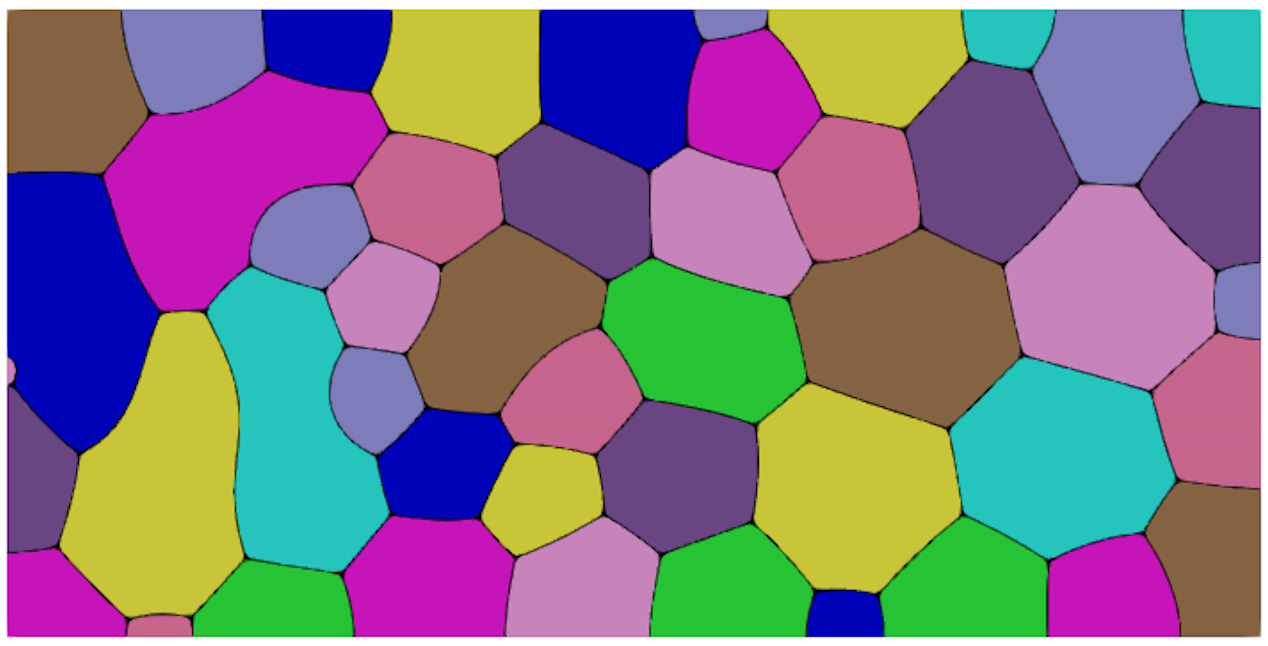}
  }
  \caption{Grain structure for $M_i  = 4.25 \times10^{17} \ \upmu\text{m}^3/(\text{J}.\text{s})$. (b)  (c): \(\Delta t = 8 \times 10^{-7}\) s, (d)  (e): \(\Delta t = 1.6 \times 10^{-6}\) s.}
  \label{Lac1}
\end{figure}

Figure~\ref{Lac1}(a) shows the reference phase field solution at \(1.6 \times 10^{-4}\) s  obtained with the mobility $M_i=4.25 \times10^{17} \ \upmu\text{m}^3/(\text{J}.\text{s})$. In this case, due to the large mobility value, the instability presents with smaller time steps. Figure~\ref{Lac1}(b)(c) show the results obtained with the first-order semi-implicit schemes with a time step size $\Delta t = 8 \times 10^{-7}$ s. The stabilized semi-implicit scheme again outperforms  the unstabilized one in terms of capturing the movement of the GBs with the given time step.  Figure~\ref{Lac1}(d)(e) confirm this observation.

\begin{figure}[htbp]
  \centering
  \includegraphics[width=0.7\textwidth]{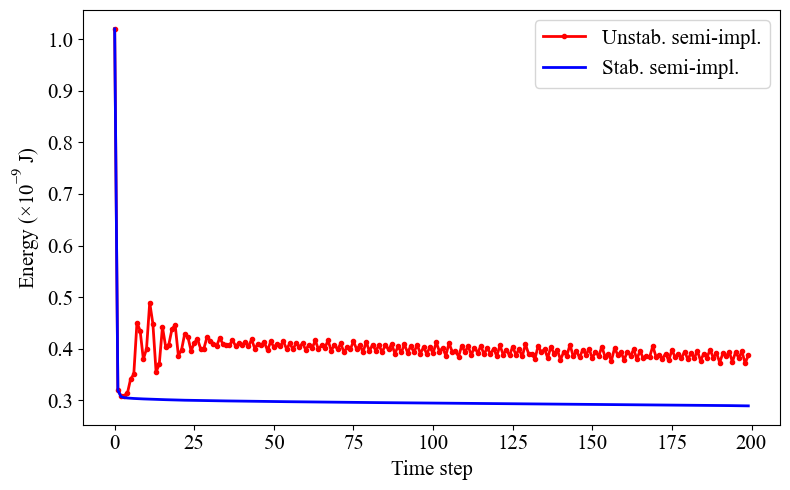}
  \caption{Energy evolution}
  \label{LbsEner}
\end{figure}

To further confirm the energy stability, we calculated the energy evolution in the above cases. Figure~\ref{LbsEner} shows two representative examples of the energy evolution for stabilized semi-implicit schemes and the unstabilized ones. It can be seen that the energy produced by the unstabilized scheme is not monotonically decreasing and shows strong oscillations over time. In contrast, the stabilized semi-implicit scheme can strictly ensure the decrease of energy over time, which is expected from our theoretical analysis.

This numerical study confirms the energy stability of the proposed stabilized semi-implicit formulation  under the isothermal condition for pure solid phases, which corresponds to the first part of Eq. \eqref{eq:energylaw_split} for $\Omega_{ss}$. To account for the solid-liquid interface energy, i.e., the second part of Eq. \eqref{eq:energylaw_split}, we would need to conduct a full solidification simulation, which is presented in the next.  


\subsection{2D grain growth simulations}
\begin{figure}[htbp]
  \centering
  \hspace*{-0.02\textwidth}
  \captionsetup[subfigure]{labelformat=empty}
  \subcaptionbox{\hspace{2em}(a)\, $t=0 \, \upmu $s}[0.35\textwidth]{%
    \includegraphics[width=\linewidth]{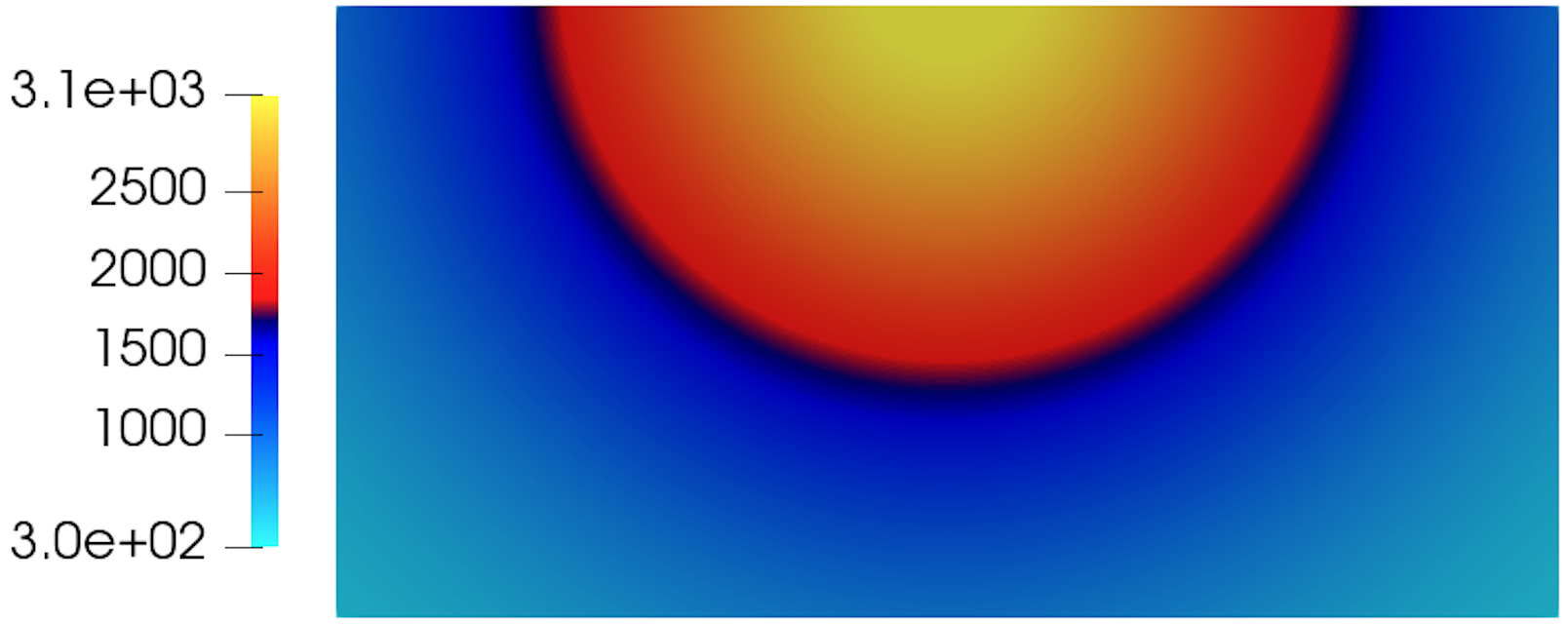}
  }
  \hspace{0.03\textwidth}
  \subcaptionbox{(b)\, $t=30 \, \upmu $s}[0.28\textwidth]{%
    \includegraphics[width=\linewidth]{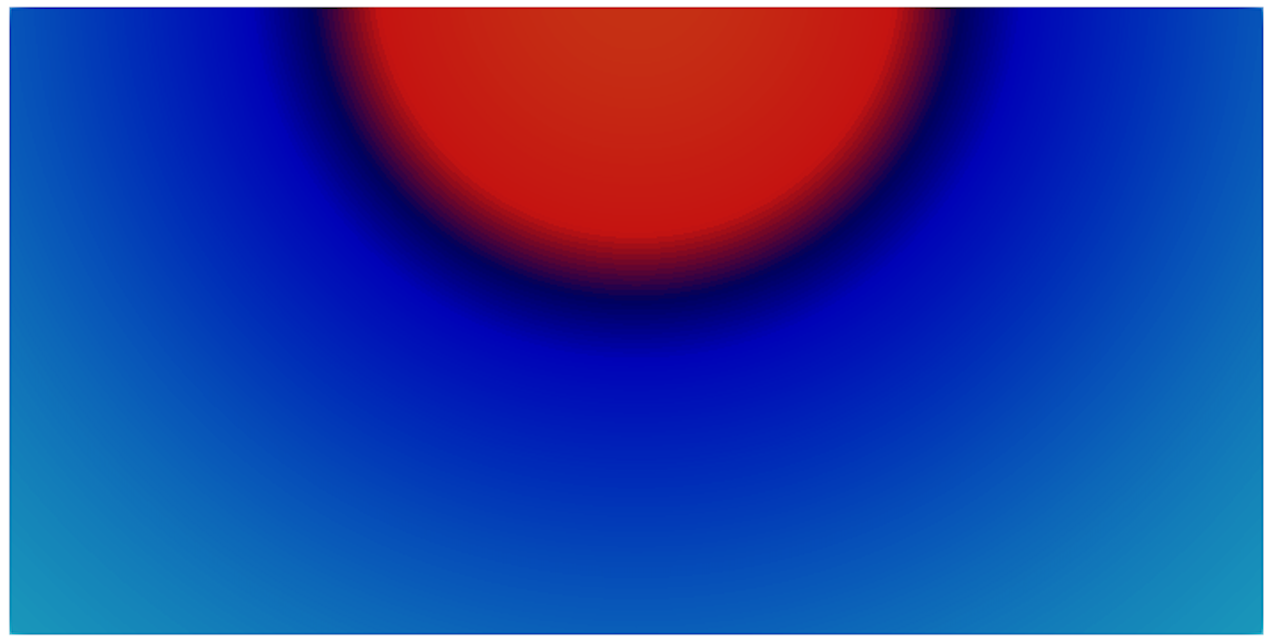}
  }
  \hspace{0.03\textwidth}
  \subcaptionbox{(c)\, $t=60 \, \upmu $s}[0.28\textwidth]{%
    \includegraphics[width=\linewidth]{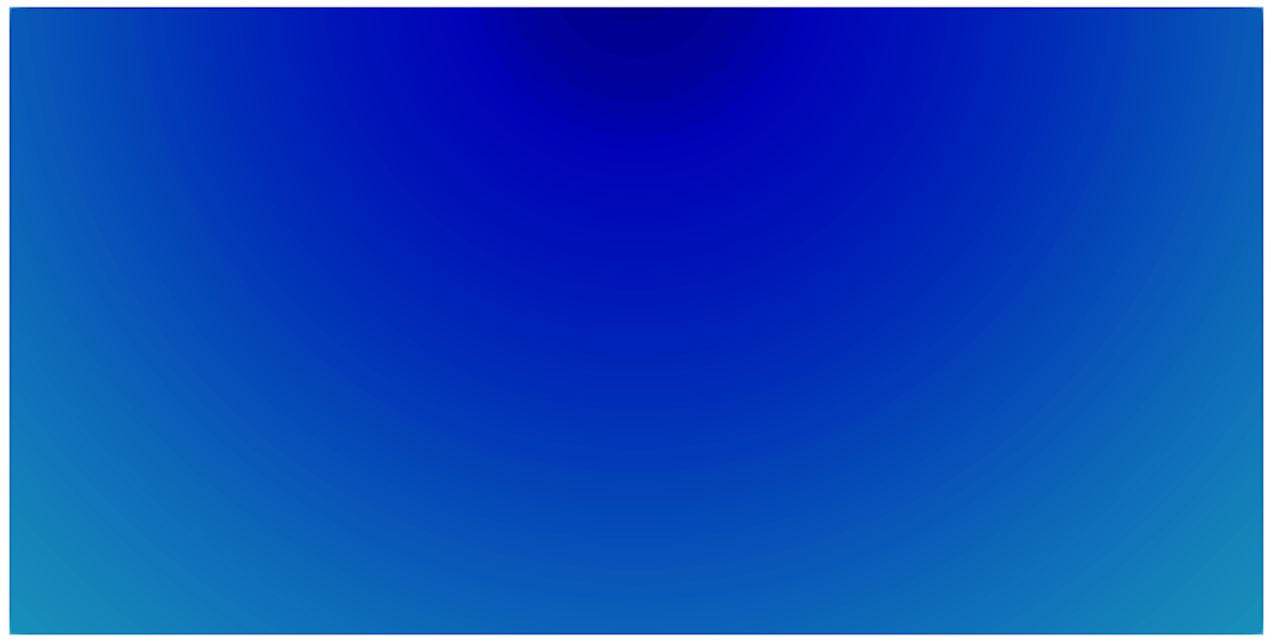}
  }  \\
     \vspace{5pt}
  \hspace*{0.02\textwidth}
  \subcaptionbox{(d)\, $t=0 \, \upmu $s}[0.312\textwidth]{%
    \includegraphics[width=\linewidth]{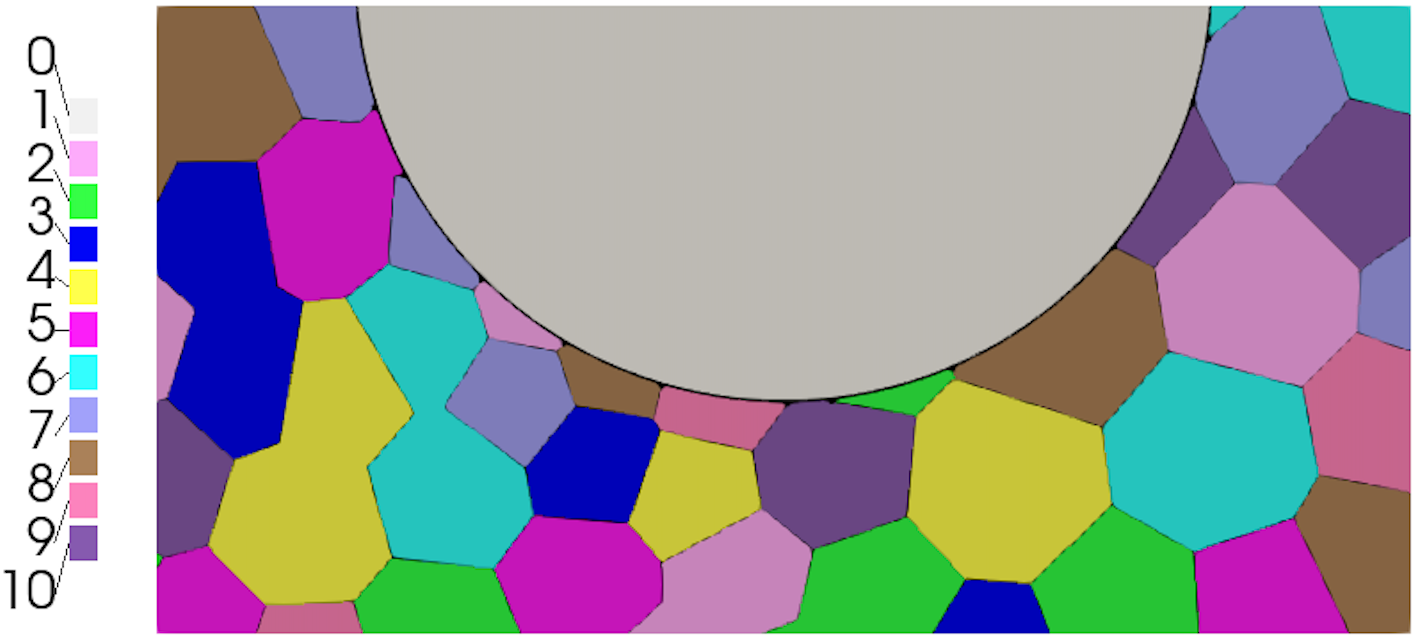}
  }
  \hspace{0.03\textwidth}
  \subcaptionbox{(e)\, $t=30 \, \upmu $s}[0.28\textwidth]{%
    \includegraphics[width=\linewidth]{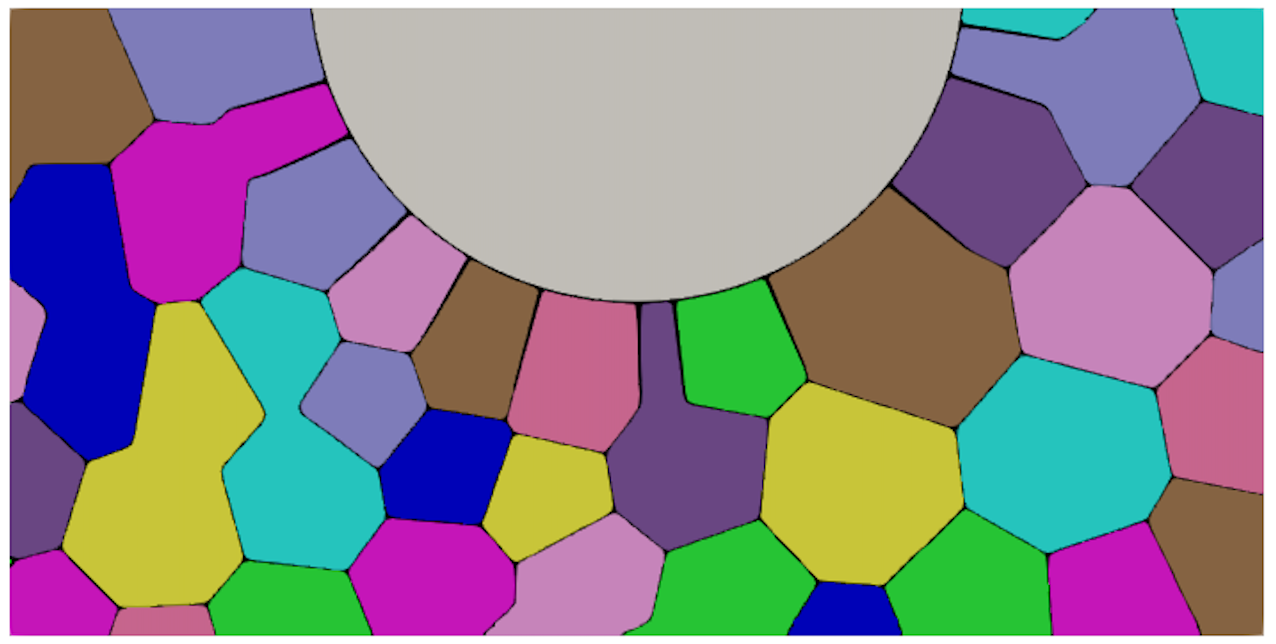}
  }
  \hspace{0.03\textwidth}
  \subcaptionbox{(f)\, $t=60 \, \upmu $s}[0.28\textwidth]{%
    \includegraphics[width=\linewidth]{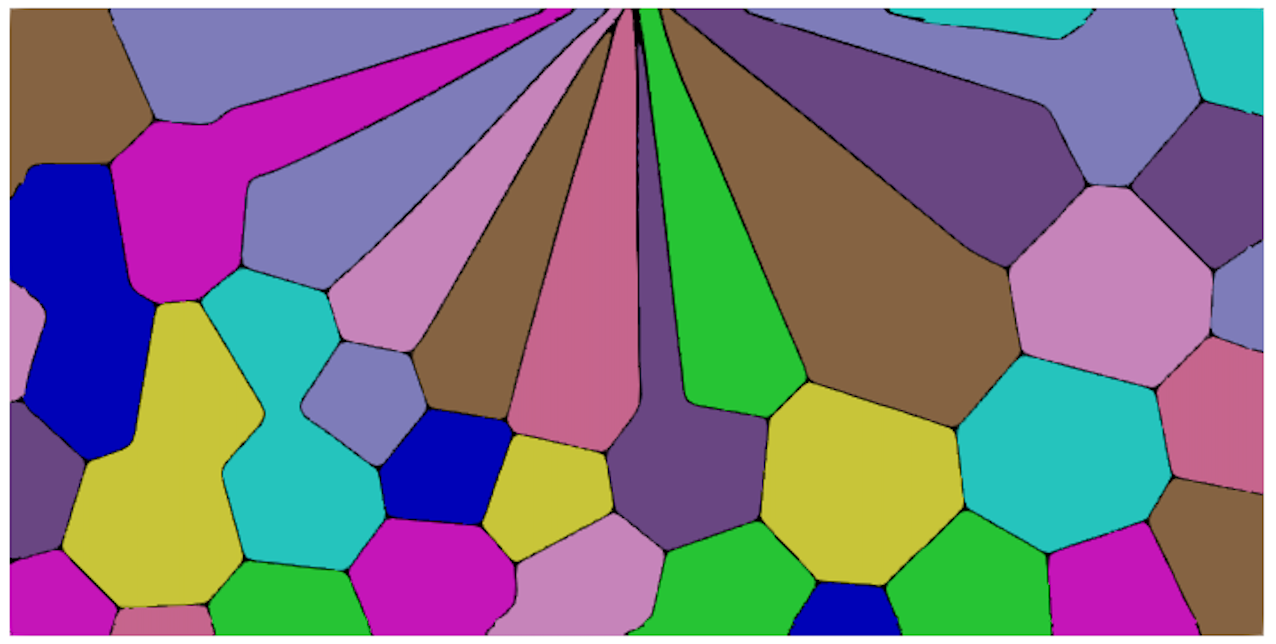}
  }    
  \caption{Reference temperature field and grain structure at different time instances}
  \label{ref42}
\end{figure}

We present here the 2D grain growth simulations. The material properties and the problem setup remain the same as the previous cases with pure solid phases, except that a liquid phase $\phi_0$ is included, as shown in \figurename~\ref{ref42}. All the phase field variables $\phi_i$, i.e., OPs, evolve with Eq. \eqref{eq:govern} under a the given temperature evolution.  {There are in total 11 OPs, including all the solid and liquid phases.}  The reference results in \figurename~\ref{ref42} are obtained by the second-order semi-implicit time integration scheme with a very small time step $5\times10^{-9}$ s. We can see that the epitaxial grain growth follows the negative gradient direction of temperature at a rapid solidification rate, which is consistent with our expectation.

\begin{figure}[htbp]
  \centering
  \subcaptionbox{\,\(\Delta t = 5 \times 10^{-9}\) s }[0.28\textwidth]{%
    \includegraphics[width=\linewidth]{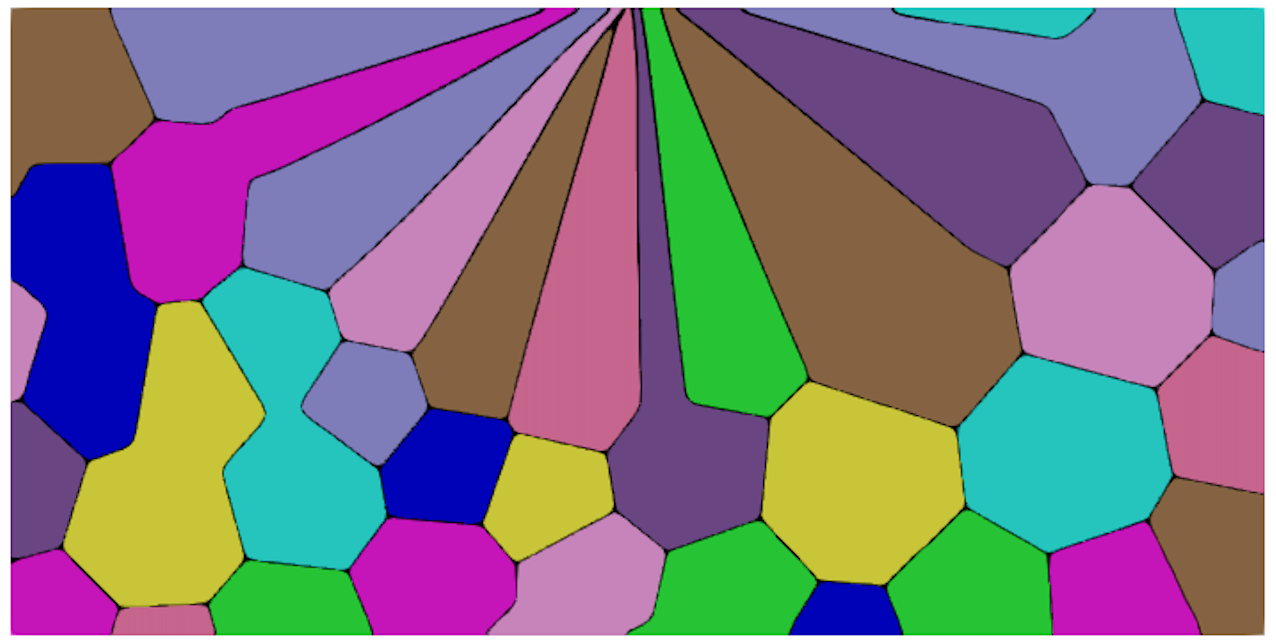}
  }
  \hspace{0.03\textwidth}
  \subcaptionbox{\,\(\Delta t = 1 \times 10^{-8}\) s}[0.28\textwidth]{%
    \includegraphics[width=\linewidth]{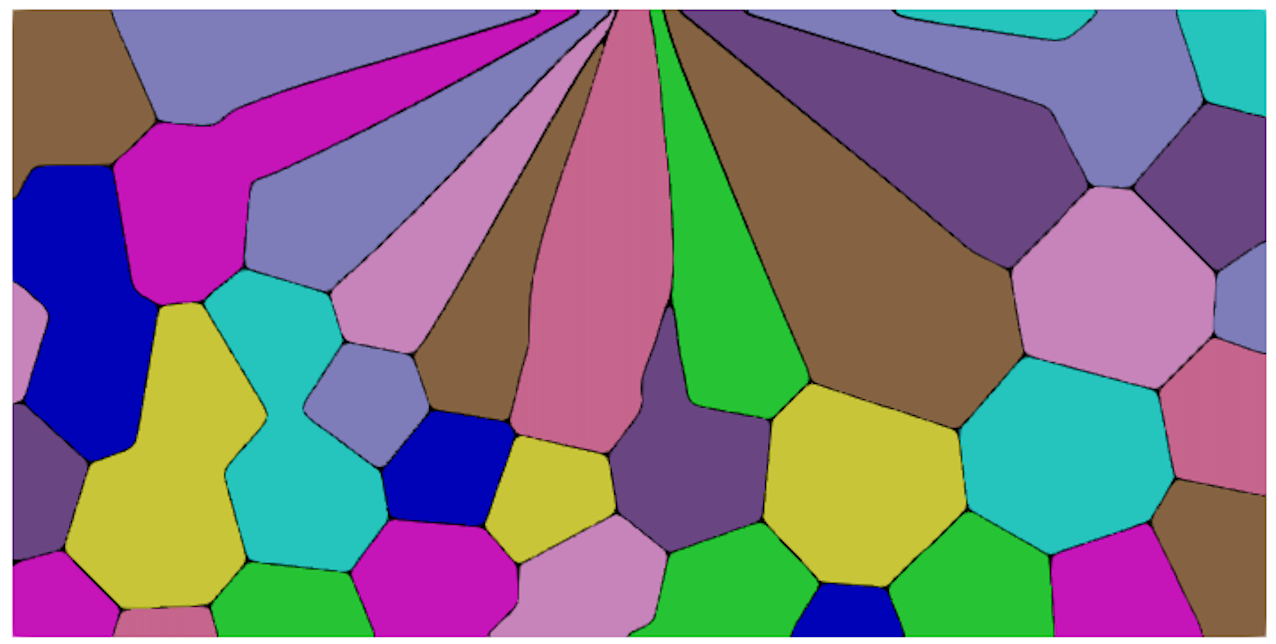}
  }
  \hspace{0.03\textwidth}
  \subcaptionbox{\,\(\Delta t = 2 \times 10^{-8}\) s}[0.28\textwidth]{%
    \includegraphics[width=\linewidth]{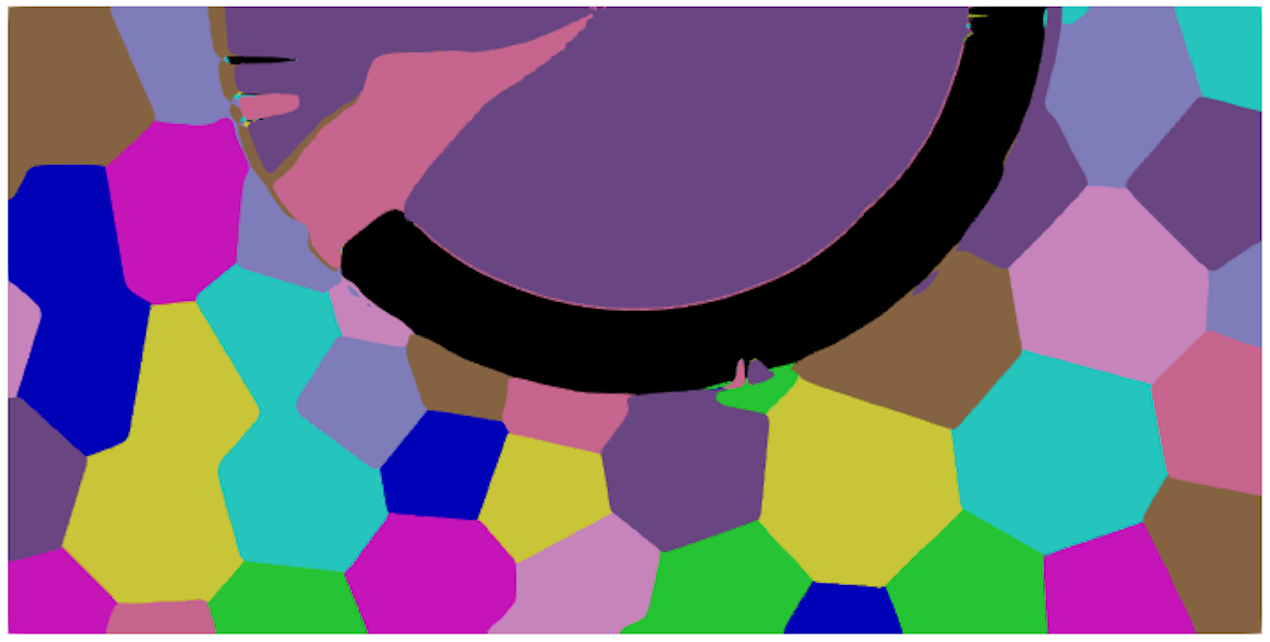}
  }
  \caption{Final grain structures obtained by the explicit scheme}
  \label{ex}
\end{figure}

\begin{figure}[htbp]
  \centering
  \subcaptionbox{\,\(\Delta t = 5 \times 10^{-9}\) s }[0.28\textwidth]{%
    \includegraphics[width=\linewidth]{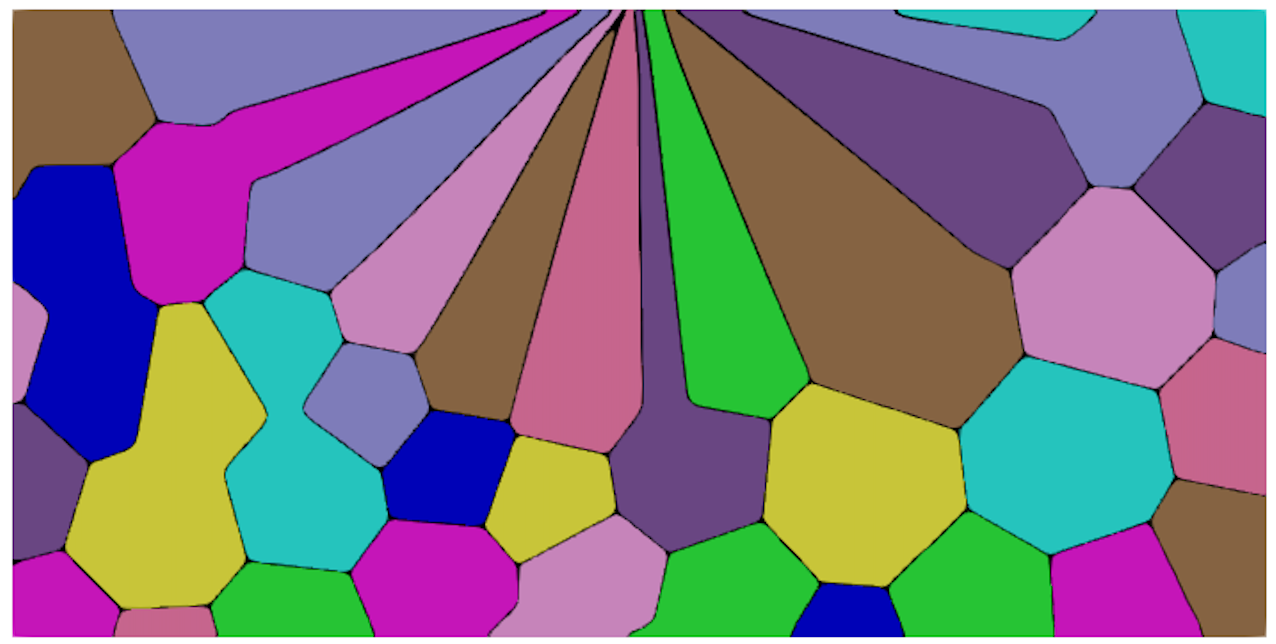}
  }
  \hspace{0.03\textwidth}
  \subcaptionbox{\,\(\Delta t = 1 \times 10^{-8}\) s}[0.28\textwidth]{%
    \includegraphics[width=\linewidth]{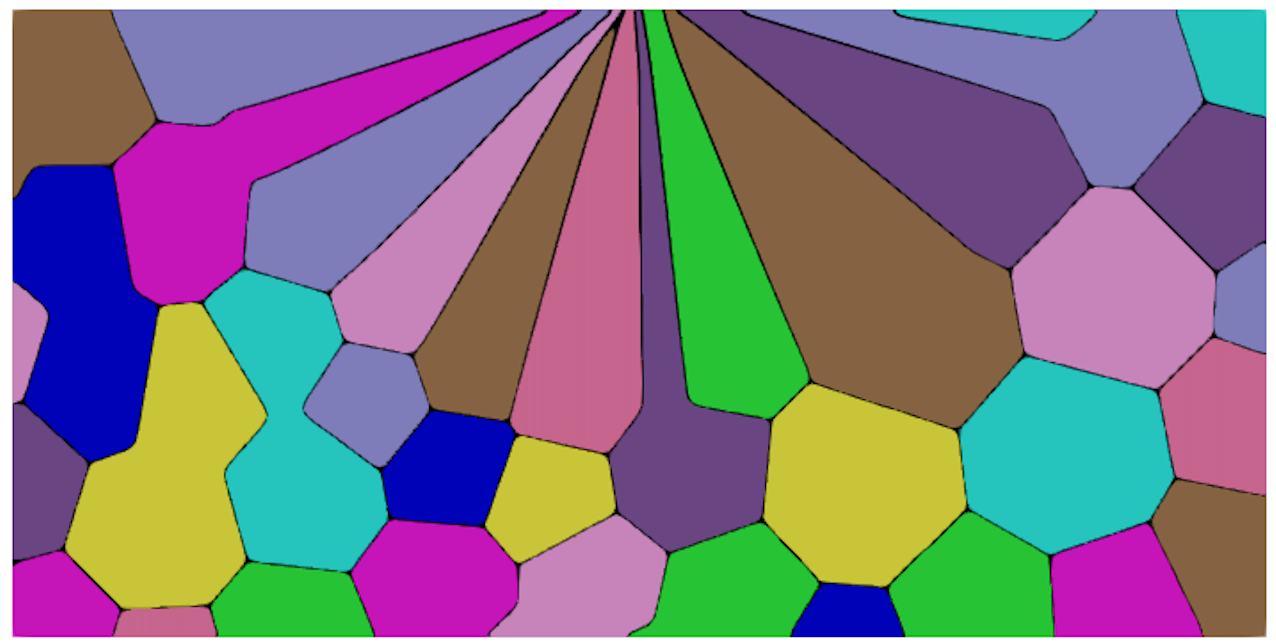}
  }
  \hspace{0.03\textwidth}
  \subcaptionbox{\,\(\Delta t = 1 \times 10^{-7}\) s}[0.28\textwidth]{%
    \includegraphics[width=\linewidth]{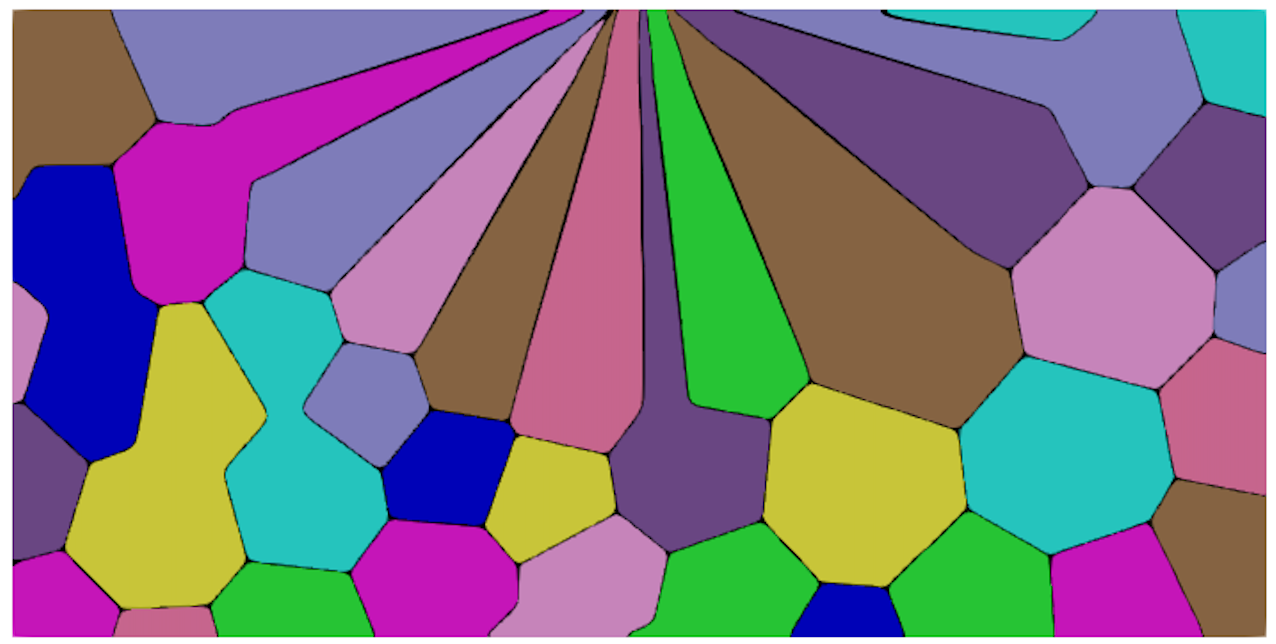}
  }
  \caption{Final grain structures obtained by the stabilized first-order semi-implicit scheme}
  \label{alpha1}
\end{figure}

\begin{figure}[htbp]
  \centering
  \subcaptionbox{\,\(\Delta t = 5 \times 10^{-9}\) s }[0.28\textwidth]{%
    \includegraphics[width=\linewidth]{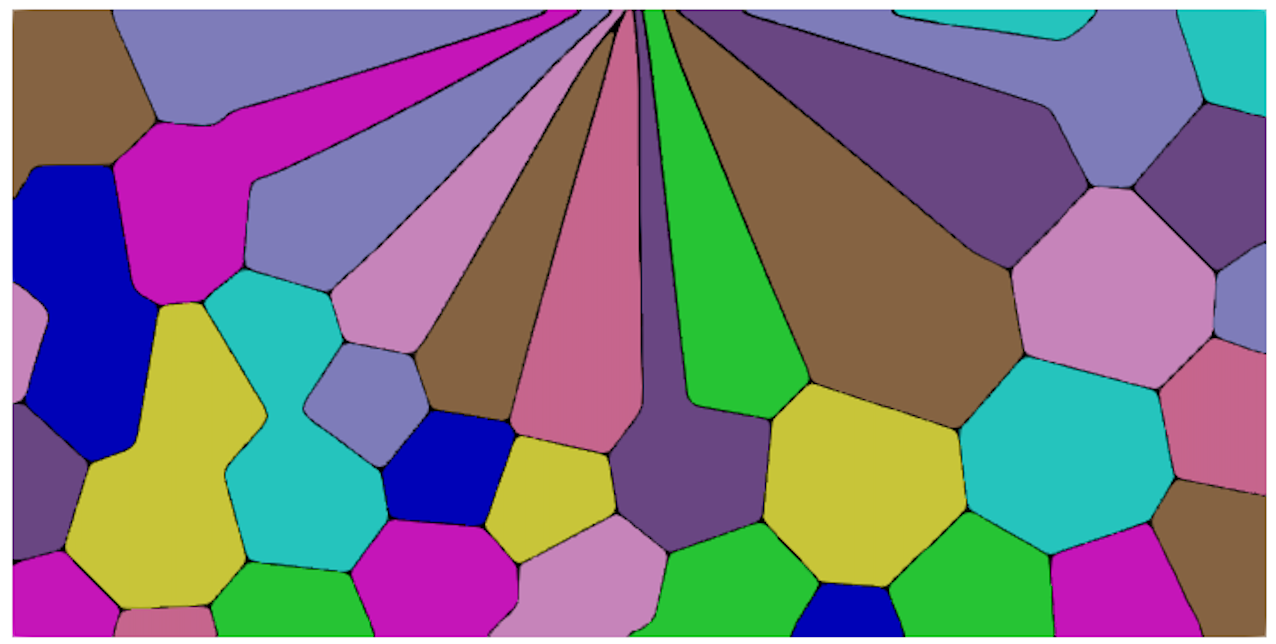}
  }
  \hspace{0.03\textwidth}
  \subcaptionbox{\,\(\Delta t = 1 \times 10^{-8}\) s}[0.28\textwidth]{%
    \includegraphics[width=\linewidth]{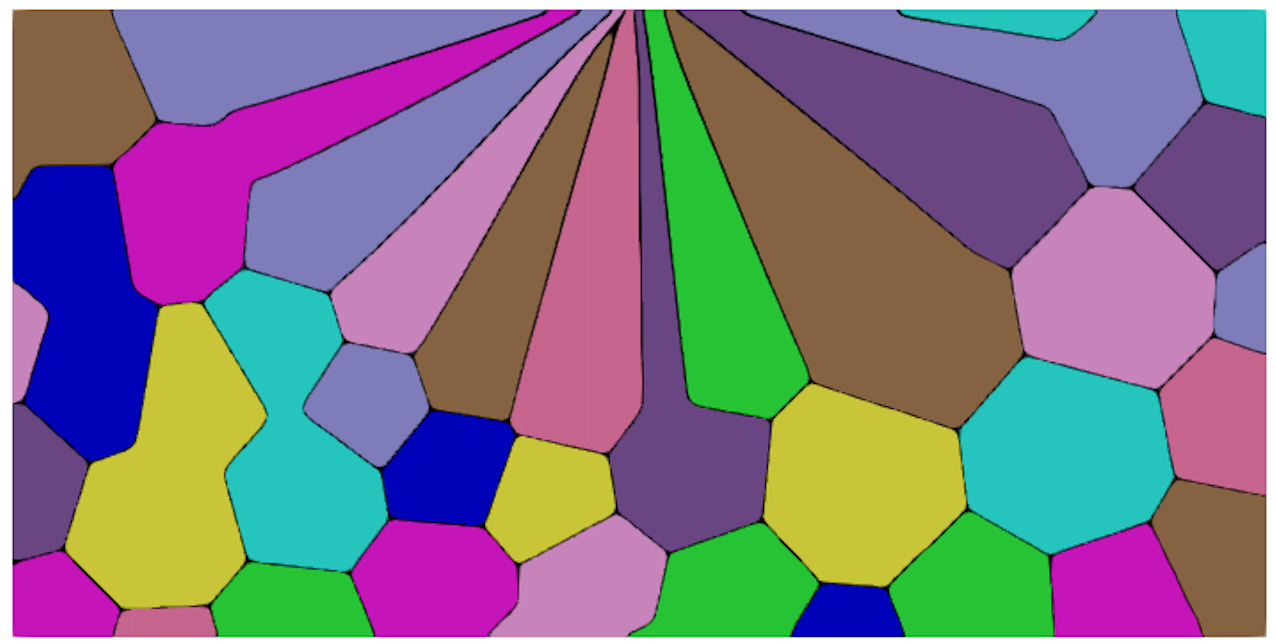}
  }
  \hspace{0.03\textwidth}
  \subcaptionbox{\,\(\Delta t = 1 \times 10^{-7}\) s}[0.28\textwidth]{%
    \includegraphics[width=\linewidth]{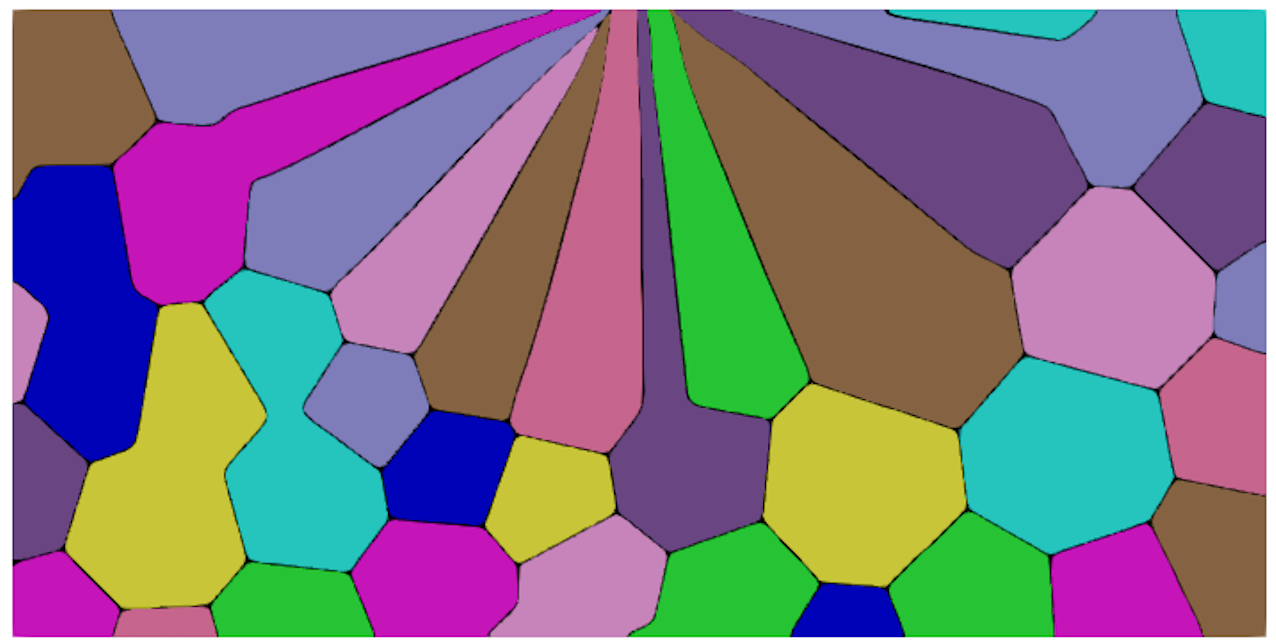}
  }
  \caption{Final grain structures obtained by the stabilized second-order semi-implicit scheme}
  \label{alpha2}
\end{figure}

We compared the performance of the proposed stabilized semi-implicit schemes with the explicit Euler scheme under larger time steps. \figurename~\ref{ex} shows the results of the explicit method for three different time steps: $\Delta t = 5 \times 10^{-9}, 1 \times 10^{-8}, 2 \times 10^{-8}$ s. As we can see, the explicit scheme cannot capture the accurate grain growth behavior with a time step equal or larger than $1 \times 10^{-8}$. In general, a time step of $O(10^{-9})$ is required to capture such rapid solidification phenomena under the framework of explicit time integration. This is consistent with that reported in the literature \cite{Chadwick2021,Chadwick2025}. As a comparison, we can see that the stabilized semi-implicit schemes remain robust with large steps up to $O(10^{-7})$, which is two orders-of-magnitude larger than the explicit scheme, as shown in \figurename~\ref{alpha1} and \figurename~\ref{alpha2}. This implies a significant saving in terms of computational costs. 

Additionally, we found that the explicit scheme would require an initialization stage to obtain diffuse interfaces between the different phases, before running the solidification simulation. This initialization stage can be performed by holding the initial temperature for $1~\upmu$s using a much smaller time step than $O(10^{-9})$, if the explicit scheme is also used for initialization. Alternatively, if a semi-implicit  scheme is used for initialization, lager time steps can be used. This initialization is a necessary setting for the explicit scheme, otherwise, it would not lead to the desired results. Similar initialization settings are reported by \cite{Chadwick2021} as well. However, for the proposed semi-implicit schemes, no such requirements are needed. In another word, the solidification simulation can be performed directly with the initial grain structure. This makes another advantage for the proposed semi-implicit schemes.

\begin{figure}[htbp]
  \centering
  \includegraphics[width=0.5\textwidth]{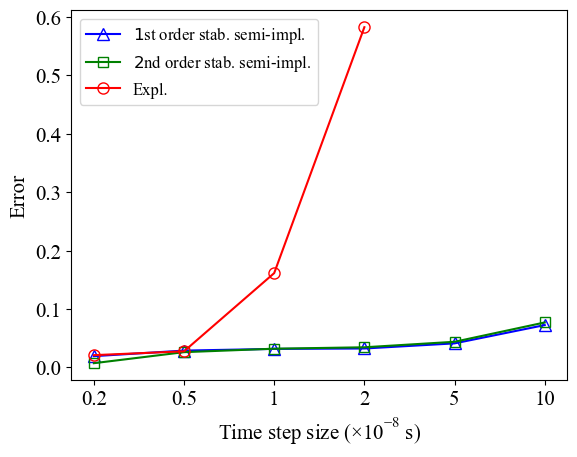}
  \caption{Error analysis}
  \label{Error}
\end{figure}

In order to quantitatively measure the accuracy of the results, we used a $L^2$ norm error of the final solution to evaluate the error of different time integration schemes. The error is defined as below
\begin{equation}
\epsilon = \frac{\sqrt{\sum_{i=1}^{N}|| \phi_{h,i}-\phi_{r, i} ||_{L^2(\Omega)}^2}}{\sqrt{\sum_{i=1}^{N}||\phi_{r, i}||_{L^2(\Omega)}^2}}
\end{equation}
where $\phi_{r, i}$ denotes the reference solution of $i$-th phase (or OP), and $\phi_{h,i}$ denotes the numerical solution obtained from a scheme being evaluated, $N$ is the total number of phases (or OPs). Figure~\ref{Error} shows the calculated error for the different schemes with different step sizes. We can see that both first-order and second-order stabilized semi-implicit schemes remain very accurate when increasing the step size, compared to the explicit one. This confirms our previous observation.

\begin{figure}[htbp]
  \centering
  \includegraphics[width=0.6\textwidth]{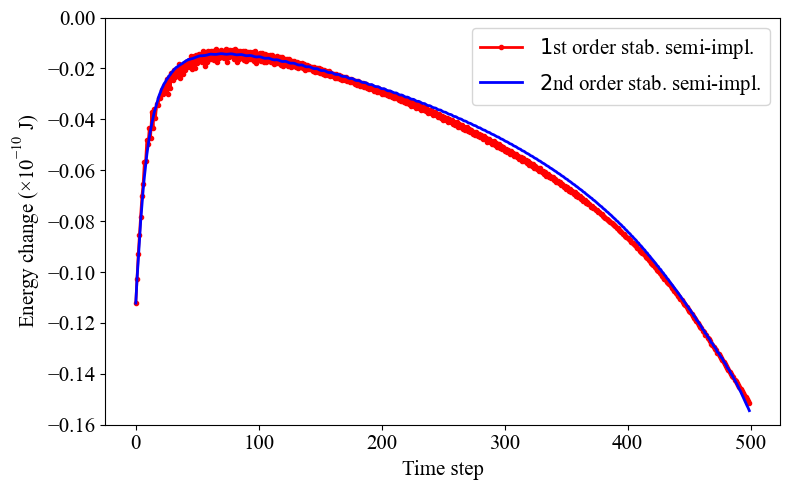}
  \caption{Energy change $\Delta E =E(\{\phi^{k+1}\},T^{k+1}) - E(\{\phi^{k}\},T^{k+1})$}
  \label{Energy change}
\end{figure}

As mentioned earlier, we also wanted to confirm that the discrete energy law \eqref{eq:energylaw} is strictly satisfied by the solutions. We therefore calculated the energy change at different time steps and plotted in \figurename~\ref{Energy change}. We can see that the energy change remains negative over time, which is consistent with the discrete energy requirement \eqref{eq:energylaw}. This complements the previous energy stability study, by accounting for the solid-liquid interface energy. Again, this is consistent with our theoretical analysis.

\begin{figure}[htbp]
  \centering
  \subcaptionbox{\,1st order unstab. semi-impl.}[0.28\textwidth]{%
    \includegraphics[width=\linewidth]{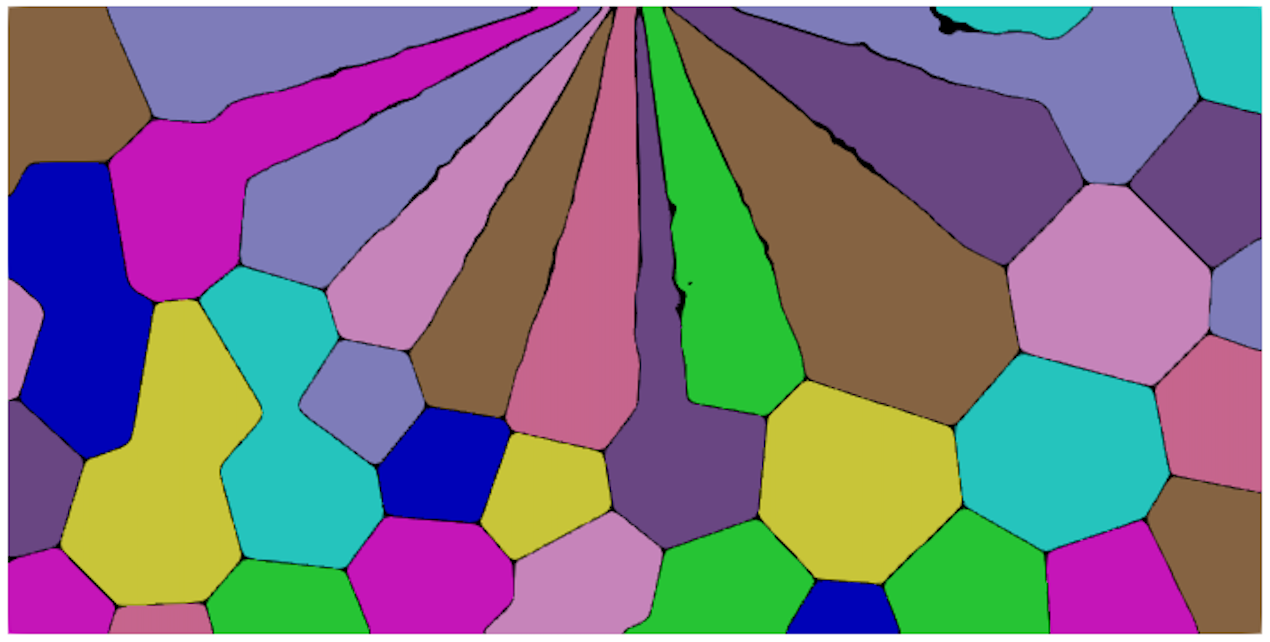}
  }
  \hspace{0.08\textwidth}
  \subcaptionbox{\,1st order stab. semi-impl.}[0.28\textwidth]{%
    \includegraphics[width=\linewidth]{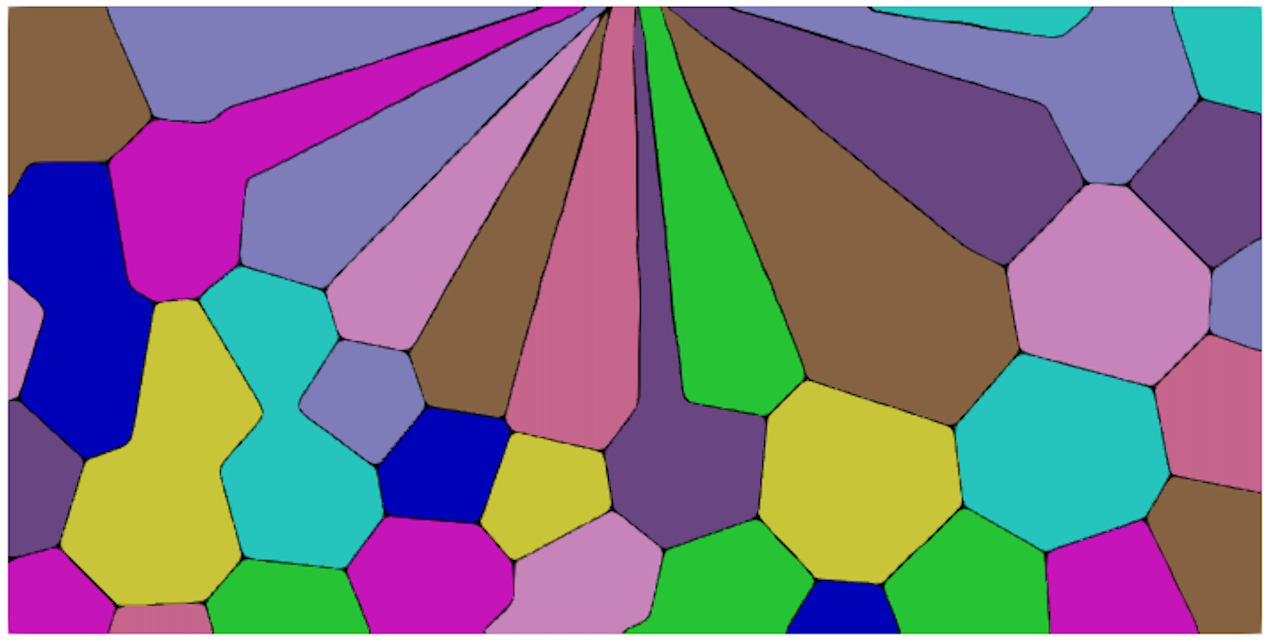}
  }
  \caption{2D grain growth simulations using a large time step: \(\Delta t = 3 \times 10^{-7}\) s}  
  \label{3-7}
\end{figure}

Finally, we wanted to confirm that the stabilized semi-implicit schemes can outperform the conventional unstabilized  schemes for the grain growth simulations. To this end, we conducted the simulations with a time step size \(3 \times 10^{-7}\,\text{s}\) and tested the first-order  stabilized semi-implicit scheme and the unstabilized one. The results are illustrated in Figure~\ref{3-7}. We can see that the unstabilized semi-implicit scheme exhibits a strong instability under such large time steps, whereas the stabilized scheme remains extremely robust. This confirms the potential of the stabilized schemes for very large time steps, which is particularly important when considering multi-time stepping algorithms. This point will be investigated in our future work. 

\subsection{3D grain growth simulations}
We now proceed to evaluate the performance of the proposed computational framework in 3D grain growth simulations. A 3D domain of size \(96\,\upmu\text{m} \times 76.8\,\upmu\text{m} \times 38.4\,\upmu\text{m}\) is used as the computational domain. The mesh is made with a grid of \(400 \times 320 \times 160\).  The characteristic length \(\zeta\) and other physical and kinetic parameters remain the same as previously.  29 grain orientation IDs are randomly assigned to the CVT generated initial grain structure.  {In this case, there are 30 OPs in total, including the liquid phase.}  We conducted the simulations with the first-order stabilized semi-implicit scheme and a time step of $O(10^{-7})$. Again, this time step is impossible for explicit schemes and, unlike explicit schemes, we do not need to conduct an initialization stage before running the solidification simulation. 

\begin{figure}[htbp]
  \centering
  \hspace{-0.07\textwidth}
  \subcaptionbox{\,$t=0 \, \upmu $s}[0.42\textwidth]{%
    \includegraphics[width=\linewidth]{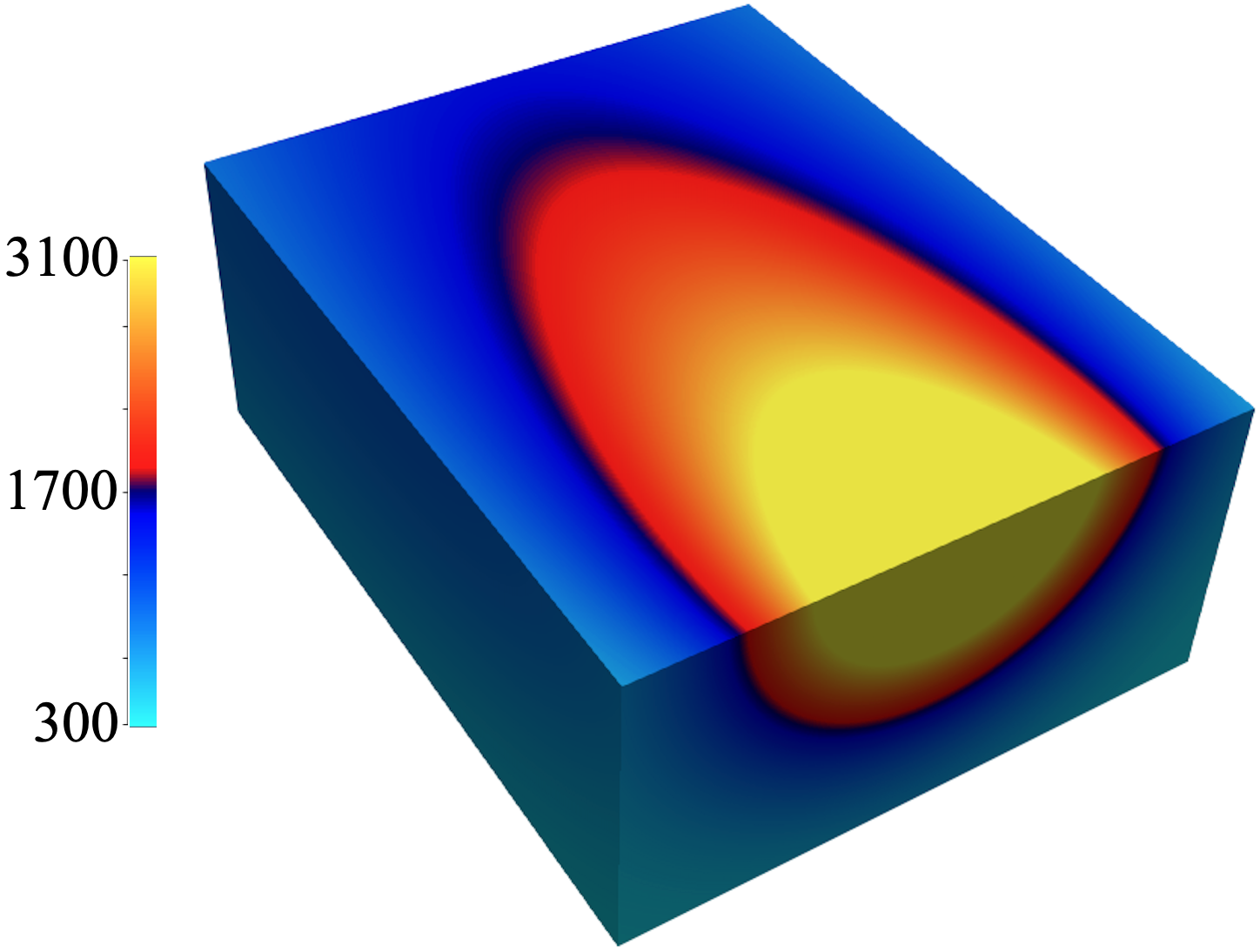}
  }
  \hspace{0.045\textwidth}
  \subcaptionbox{\,$t=0 \, \upmu $s}[0.41\textwidth]{%
    \includegraphics[width=\linewidth]{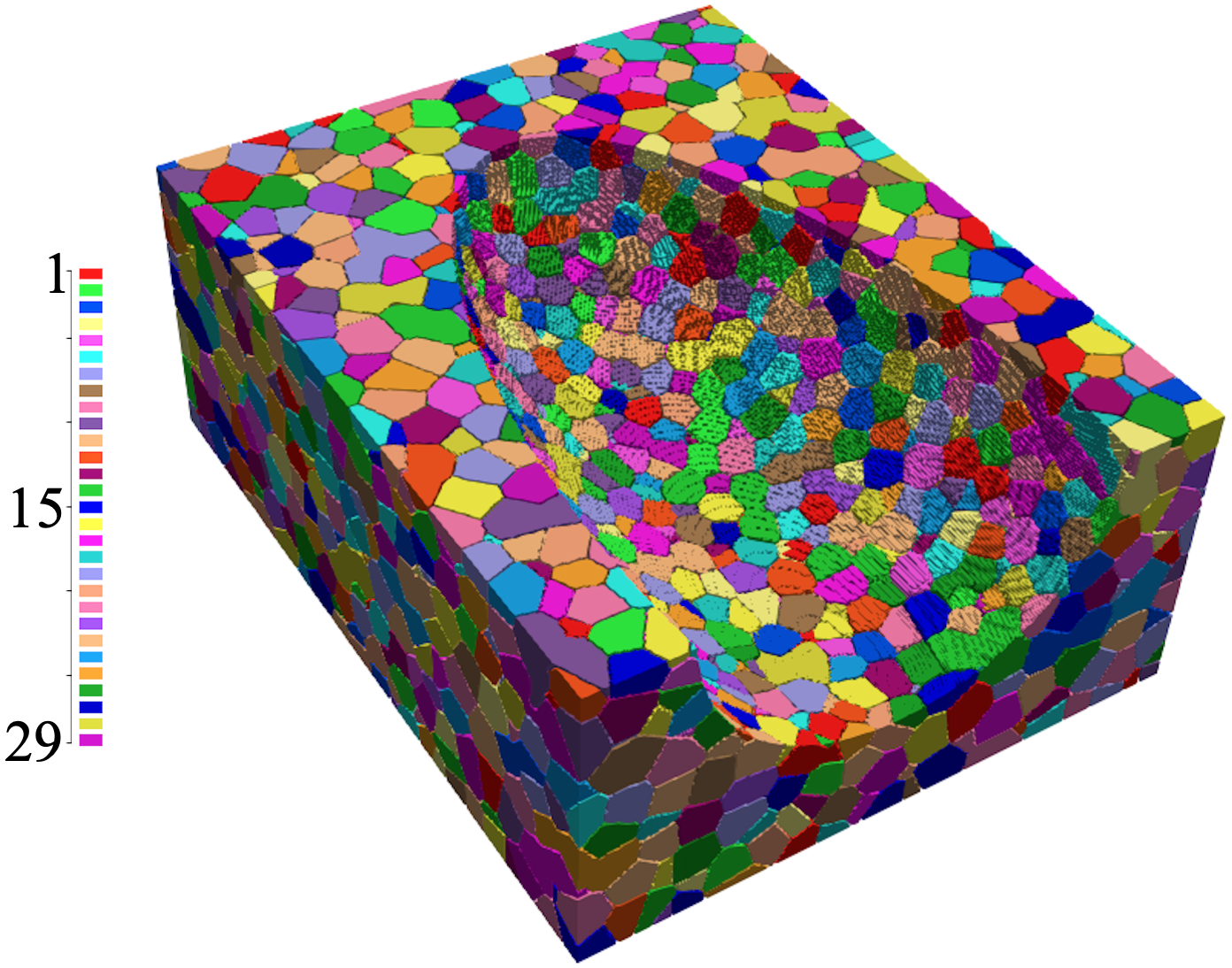}
  }\\
     \vspace{5pt}
  \hspace{0.0\textwidth}
  \subcaptionbox{\,$t=50 \, \upmu $s}[0.35\textwidth]{%
    \includegraphics[width=\linewidth]{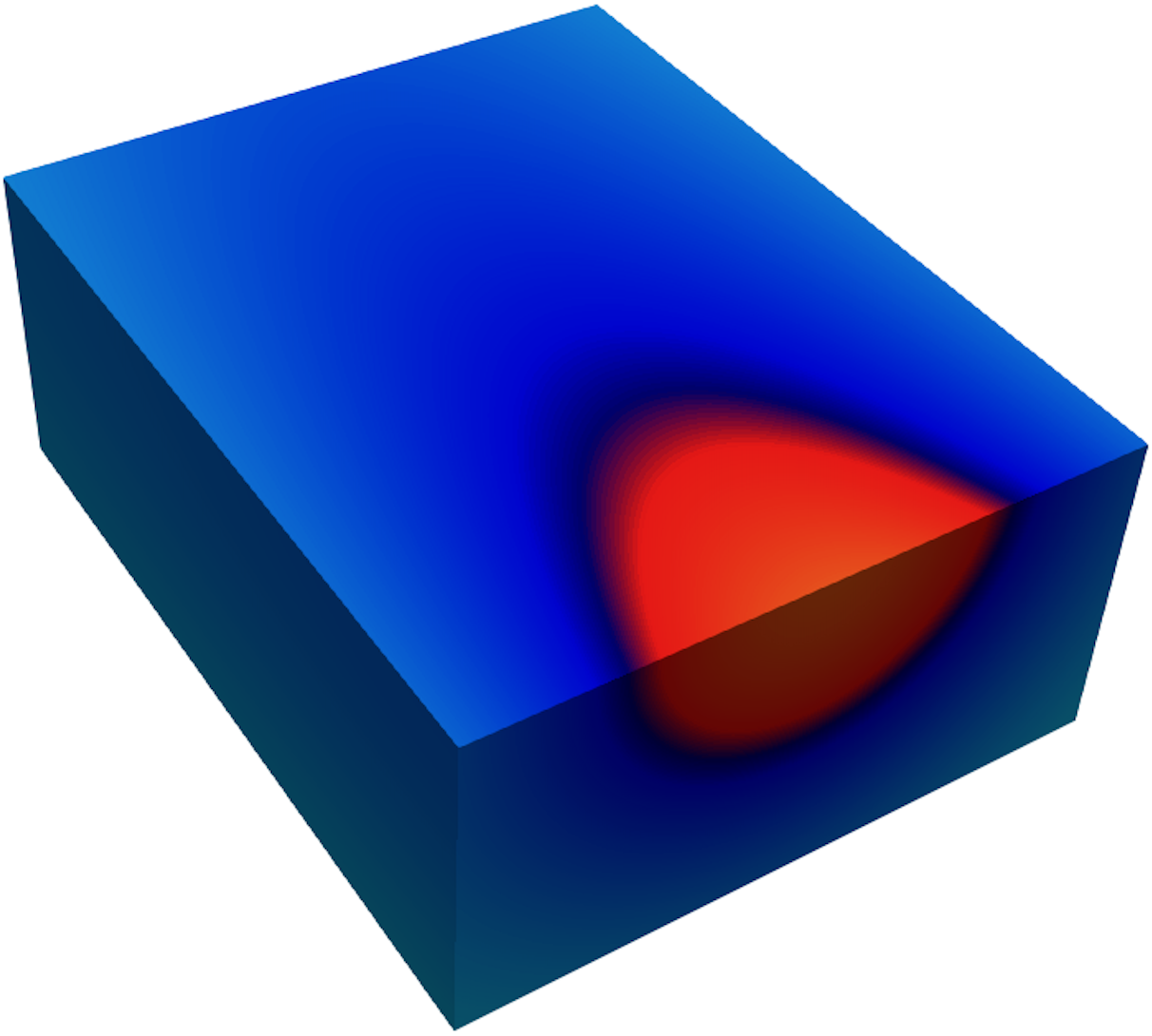}
  }  
  \hspace{0.1\textwidth}
  \subcaptionbox{\,$t=50 \, \upmu $s}[0.35\textwidth]{%
    \includegraphics[width=\linewidth]{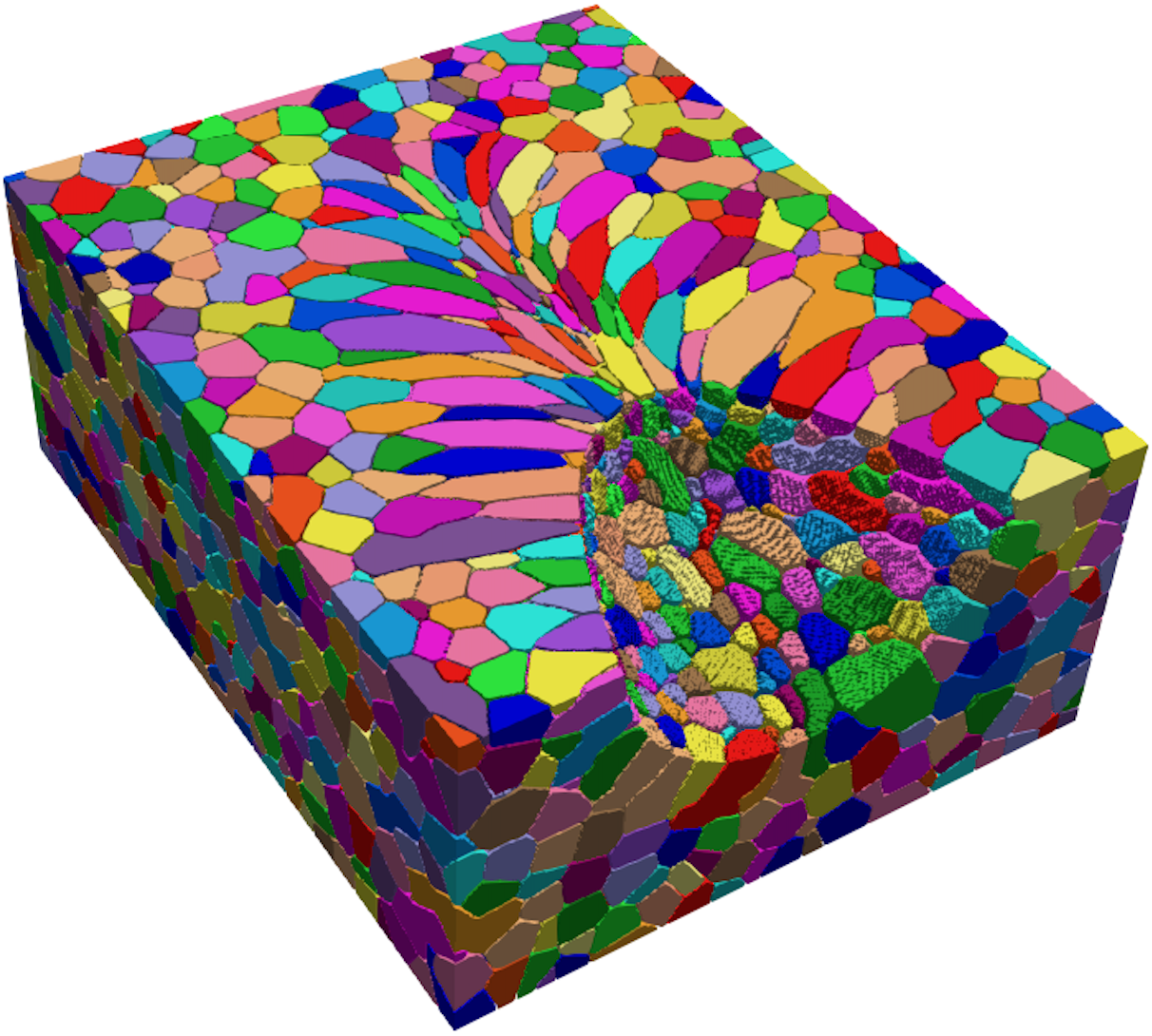}
  }\\
     \vspace{5pt}
  \hspace{0.0\textwidth}
  \subcaptionbox{\,$t=100 \, \upmu $s}[0.35\textwidth]{%
    \includegraphics[width=\linewidth]{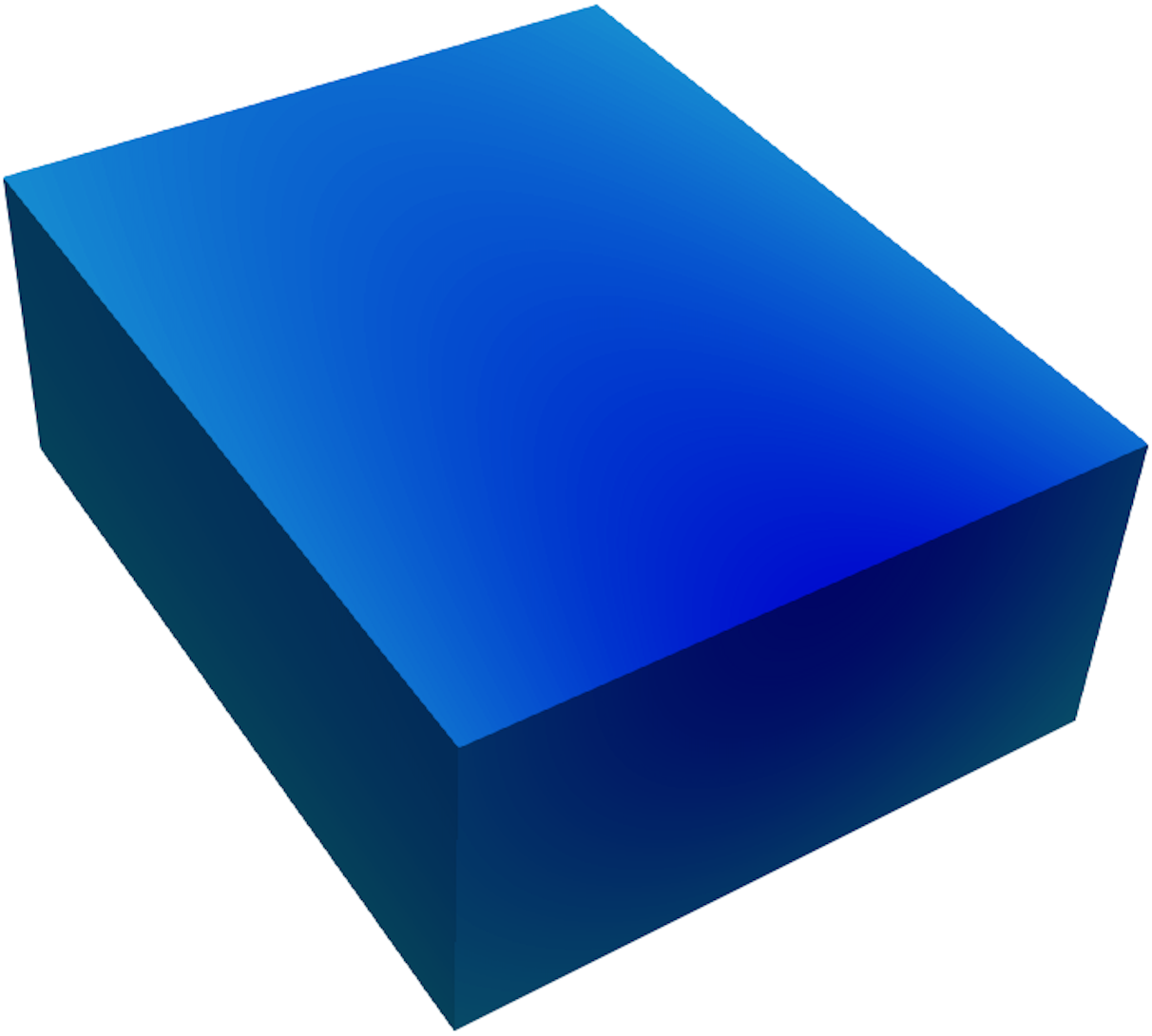}
  }
  \hspace{0.1\textwidth}
  \subcaptionbox{\, $t=100 \, \upmu $s}[0.35\textwidth]{%
    \includegraphics[width=\linewidth]{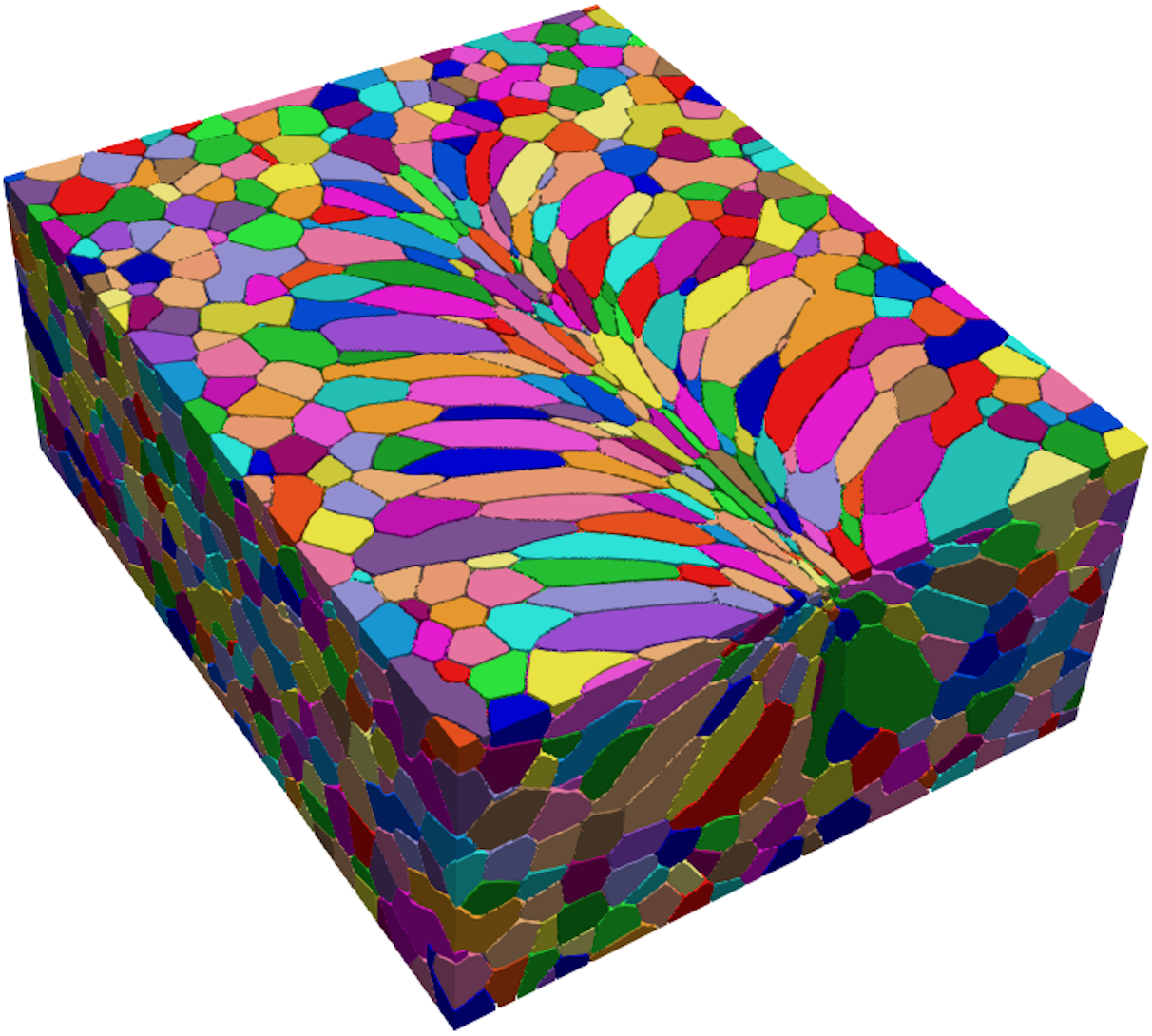}
  }    
  \caption{ {3D grain growth under the given temperature evolution at a laser scan speed $V_p=1.0 $ m/s. (a) (c) (e): temperature profiles; (b) (d) (f): grain structures}}
  \label{3d}
\end{figure}

\begin{figure}[htbp]
  \centering

  \subcaptionbox{\,$t=0 \, \upmu $s}[0.22\textwidth]{%
    \includegraphics[width=\linewidth]{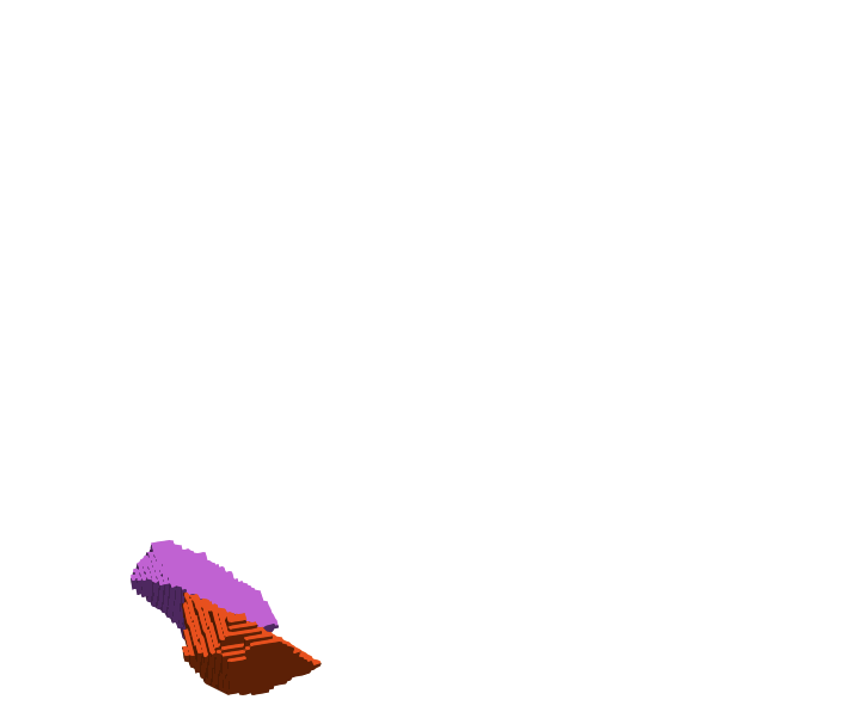}
  }
  \hspace{0.05\textwidth}
  \subcaptionbox{\,$t=15 \, \upmu $s}[0.22\textwidth]{%
    \includegraphics[width=\linewidth]{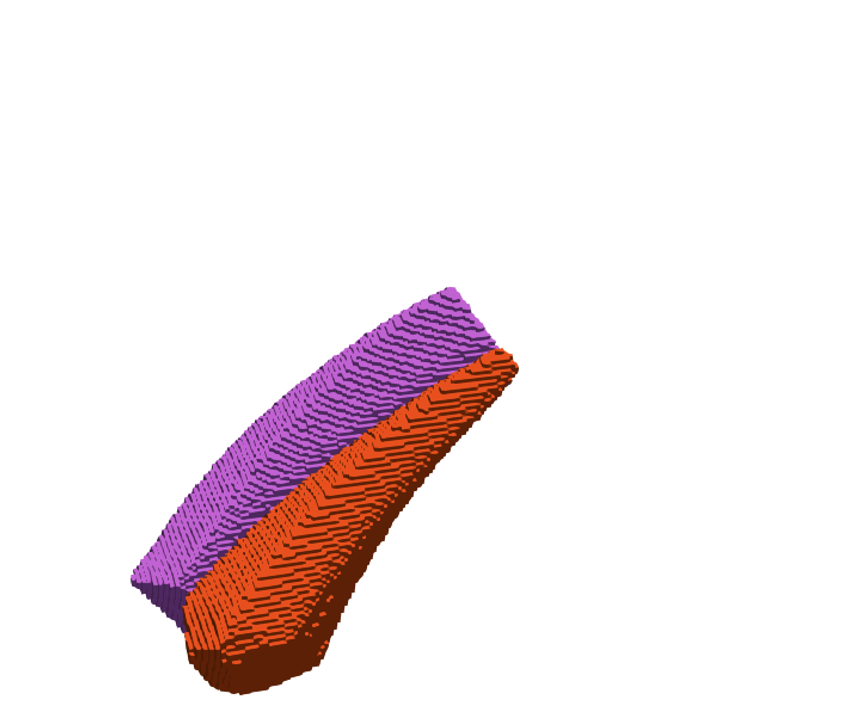}
  }
  \hspace{0.05\textwidth}
  \subcaptionbox{\,$t=30 \, \upmu $s}[0.22\textwidth]{%
    \includegraphics[width=\linewidth]{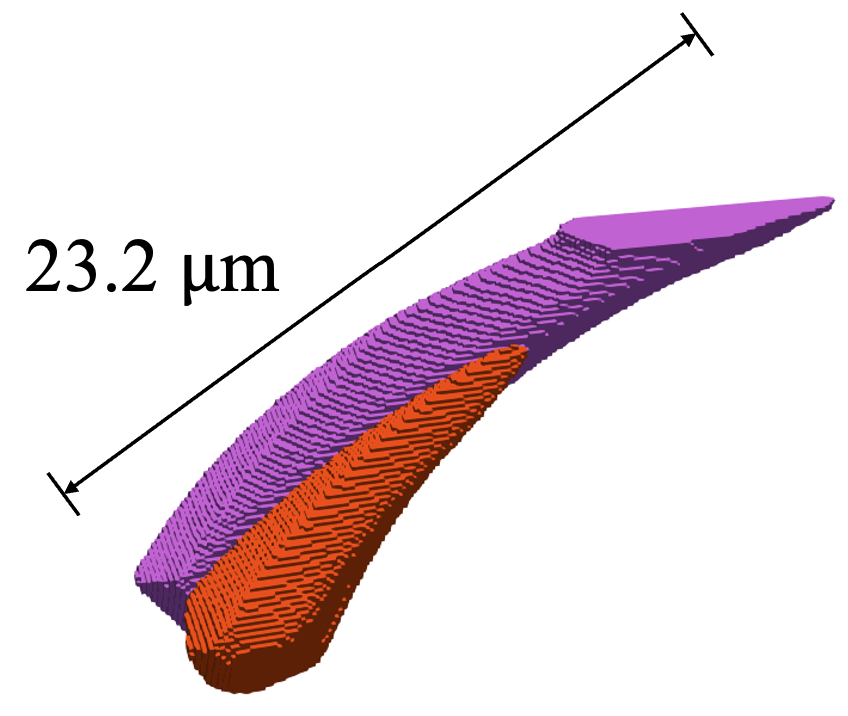}
  }  
  \caption{ {Growing shape of two competing grains at a solidification rate $\approx$ 0.7 m/s}}
  \label{3dsp}
\end{figure}

\figurename~\ref{3d} illustrates the 3D grain growth simulation results up to $100 \, \upmu$s. We can clearly see the epitaxial growth of columnar grains and their competitive behaviors during the solidification process. The final grain structure is characterized by grains curving from the sides of the melt pool toward the center of the laser track. As a representative example, \figurename~\ref{3dsp}  {shows the growing shape of two competing grains. At the beginning, the two grains grew preferentially along the negative thermal gradient direction at a similar growth rate, developing a pronounced elongated shape. As time progresses, one of the grains was promoted and occupied the space ahead of the other grain, due to the better alignment of the preferred crystallographic orientation with the evolving thermal gradient direction. At $t = 30\,\upmu$s, the grains exhibit a lower aspect ratio compared to their initial roughly equiaxed shape, along with strong directional growth and pronounced curvature}.   These observations are consistent with that reported in \cite{Chadwick2021}.
Furthermore,  {we can roughly estimate the growth rate of the two grains, which is about 0.7 m/s and seems reasonably comparable to the laser scan speed $V_p=1.0 $ m/s, indicating the rapid solidification feature of AM}. 

\begin{figure}[htbp]
  \centering
  \hspace{0\textwidth}
  \subcaptionbox{\, $\mu/\mu_0$, $\epsilon_4$ = 0, $t = 10$ $\upmu $s}[0.4\textwidth]{%
    \includegraphics[width=\linewidth]{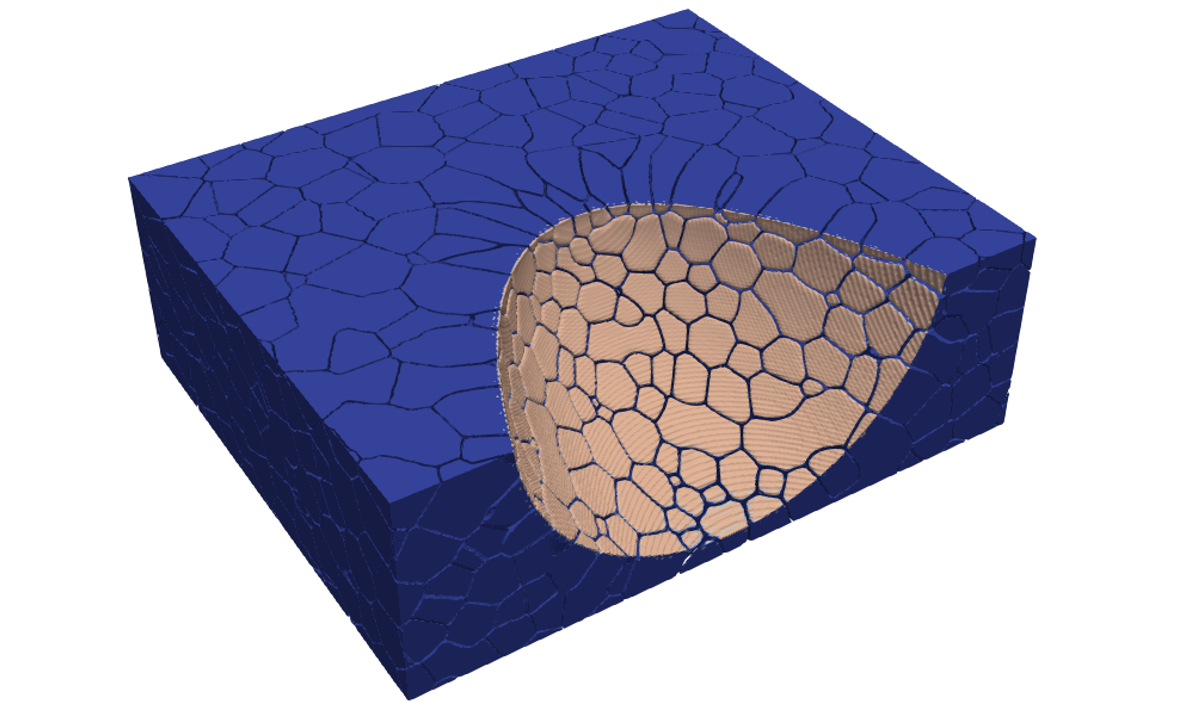}
  }
  \hspace{0\textwidth}
  \subcaptionbox{\, $\mu/\mu_0$, $\epsilon_4$ = 0.3, $t = 10$ $\upmu $s}[0.4\textwidth]{%
    \includegraphics[width=\linewidth]{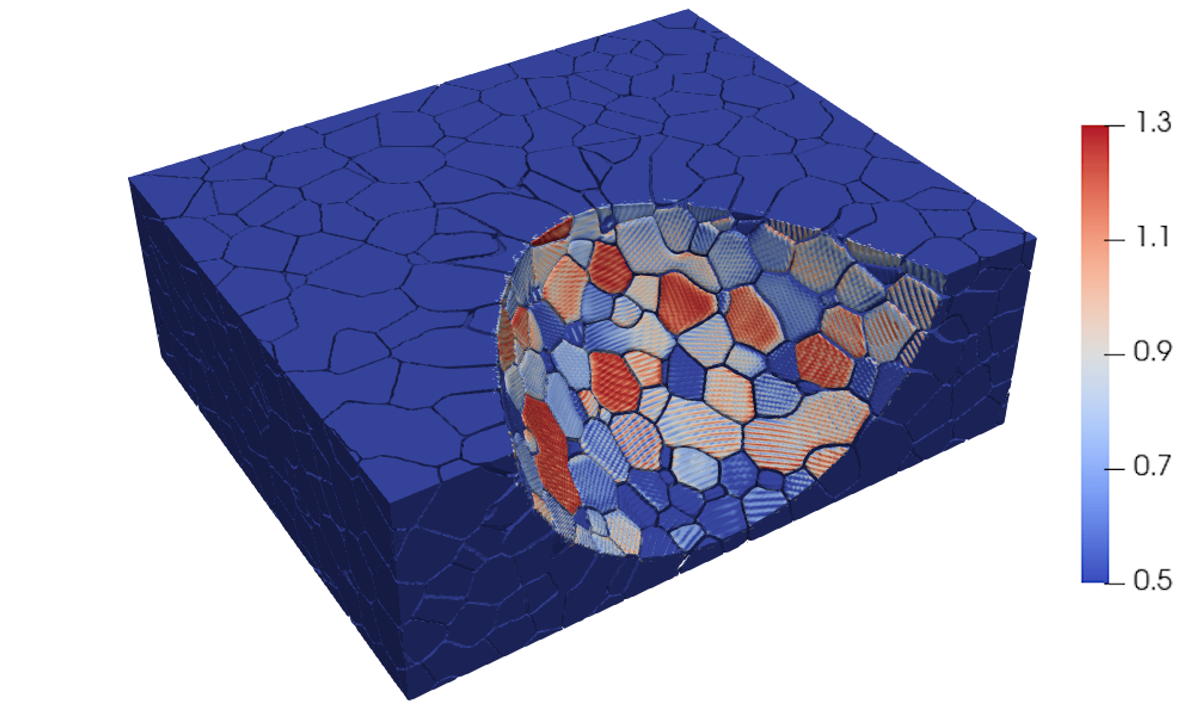}
  }
  
  \vspace{0.02\textwidth}
  \hspace{0\textwidth}
  \subcaptionbox{\, Grains, $\epsilon_4$ = 0, $t = 100$ $\upmu $s}[0.3\textwidth]{%
    \includegraphics[width=\linewidth]{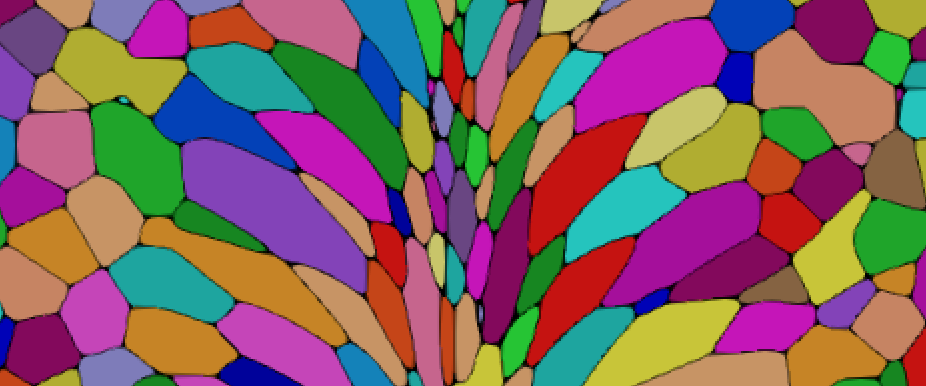}
  }
  \hspace{0.1\textwidth}
  \subcaptionbox{\, Grains, $\epsilon_4$ = 0.3, $t = 100$ $\upmu $s}[0.3\textwidth]{%
    \includegraphics[width=\linewidth]{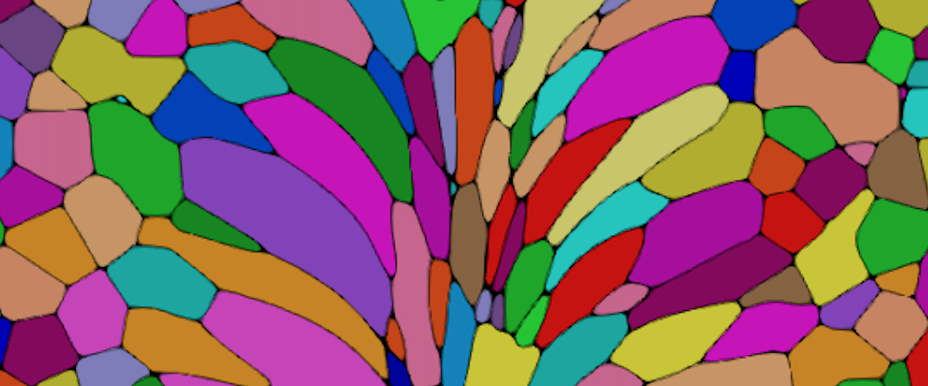}
  }

  \caption{ {Effect of kinetic anisotropy. (a) (b): Mobility coefficient $\mu/\mu_0$, (c) (d): Final grain structure, $\epsilon_4 = 0$ (isotropy) and $\epsilon_4 = 0.3$ (anisotropy)}}
  \label{kinetic anisotropy}
\end{figure}

To further validate the effectiveness of the proposed algorithm, we conducted parametric studies for two parameters: the interfacial kinetic anisotropy parameter  and the laser scan speed. We recall that the kinetic anisotropy parameter is the coefficient $\epsilon_4$ in Eq. \eqref{eq:anisoparam} and dominates the anisotropic growth behavior in the solid-liquid interfaces.  {A larger value of this parameter will promote the growth of grains having preferred crystallographic orientation most aligned with the heat flow direction}. To illustrate this effect, we rerun the 3D simulations with two distinguished values of the parameter: $\epsilon_4 = 0$ and $\epsilon_4 = 0.3$,  {as shown in \figurename~\ref{kinetic anisotropy}. \figurename~\ref{kinetic anisotropy}(a)(b) illustrate the distribution of the mobility coefficient $\mu$ among different grains on the solid-liquid interface in an early time stage.  As expected, if no anisotropic effect is included (i.e., $\epsilon_4=0$), the mobility coefficient is uniformly constant for all grains regardless of their orientations. In the case of anisotropy (i.e., $\epsilon_4=0.3$), the mobility coefficient varies significantly among different grains, according to the alignment between the preferred crystallographic growth direction and the heat flow direction. This kinetic anisotropy leads to strong grain selection and competitive growth. As shown in Fig. 14(c)(d), the isotropic case has no grain selection and all grains grew equally along the moving direction of the solid-liquid interface, whereas only a few grains were selected to grow and occupied most of the space in the melt pool region in the anisotropic case. This again agrees with the observation in \cite{Chadwick2021}}.

\begin{figure}[htbp]
  \centering
  \hspace{0\textwidth}
  \subcaptionbox{\,$V_p=1.0 $ m/s}[0.3\textwidth]{%
    \includegraphics[width=\linewidth]{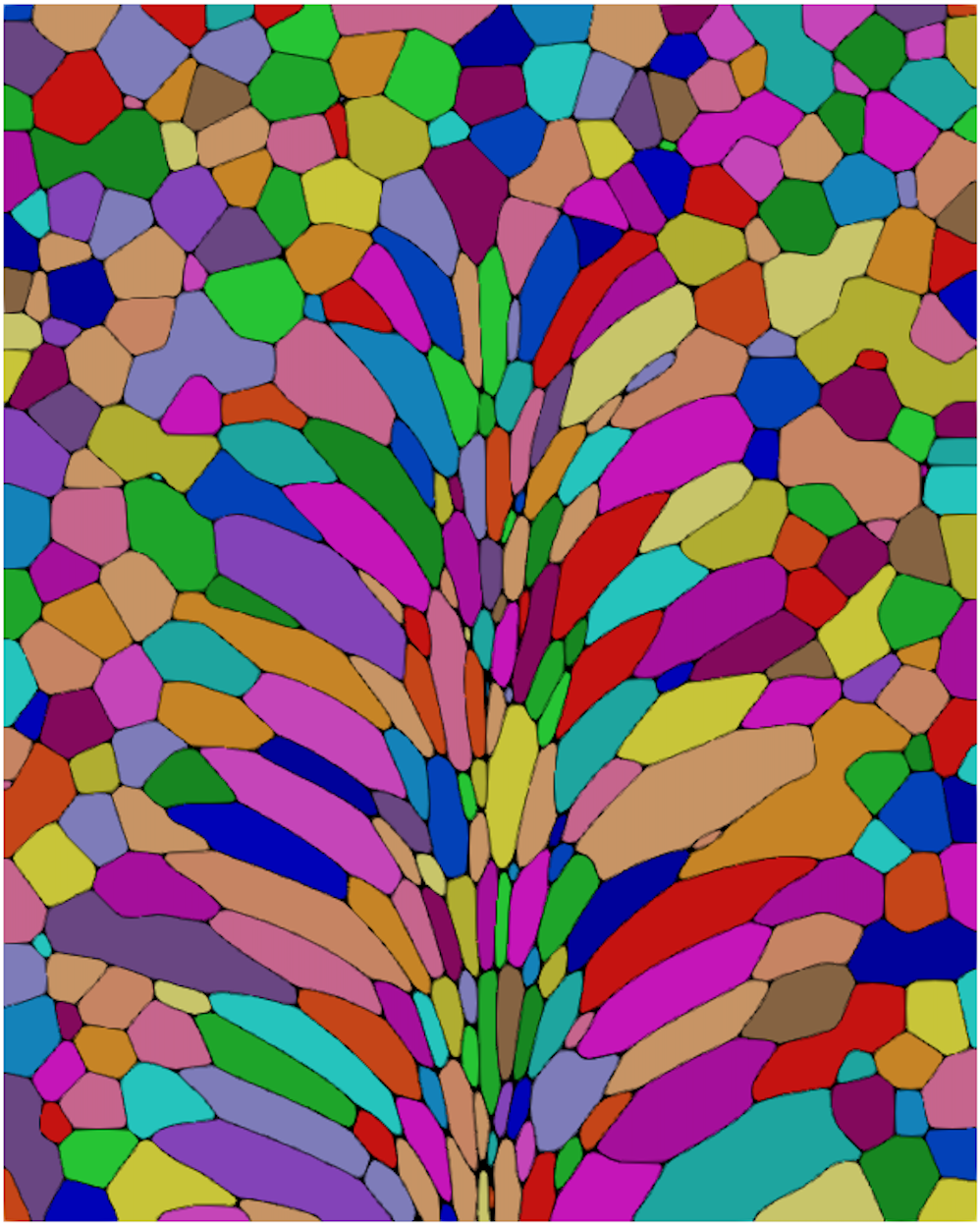}
  }
  \hspace{0.1\textwidth}
  \subcaptionbox{\,$V_p=1.5 $ m/s}[0.3\textwidth]{%
    \includegraphics[width=\linewidth]{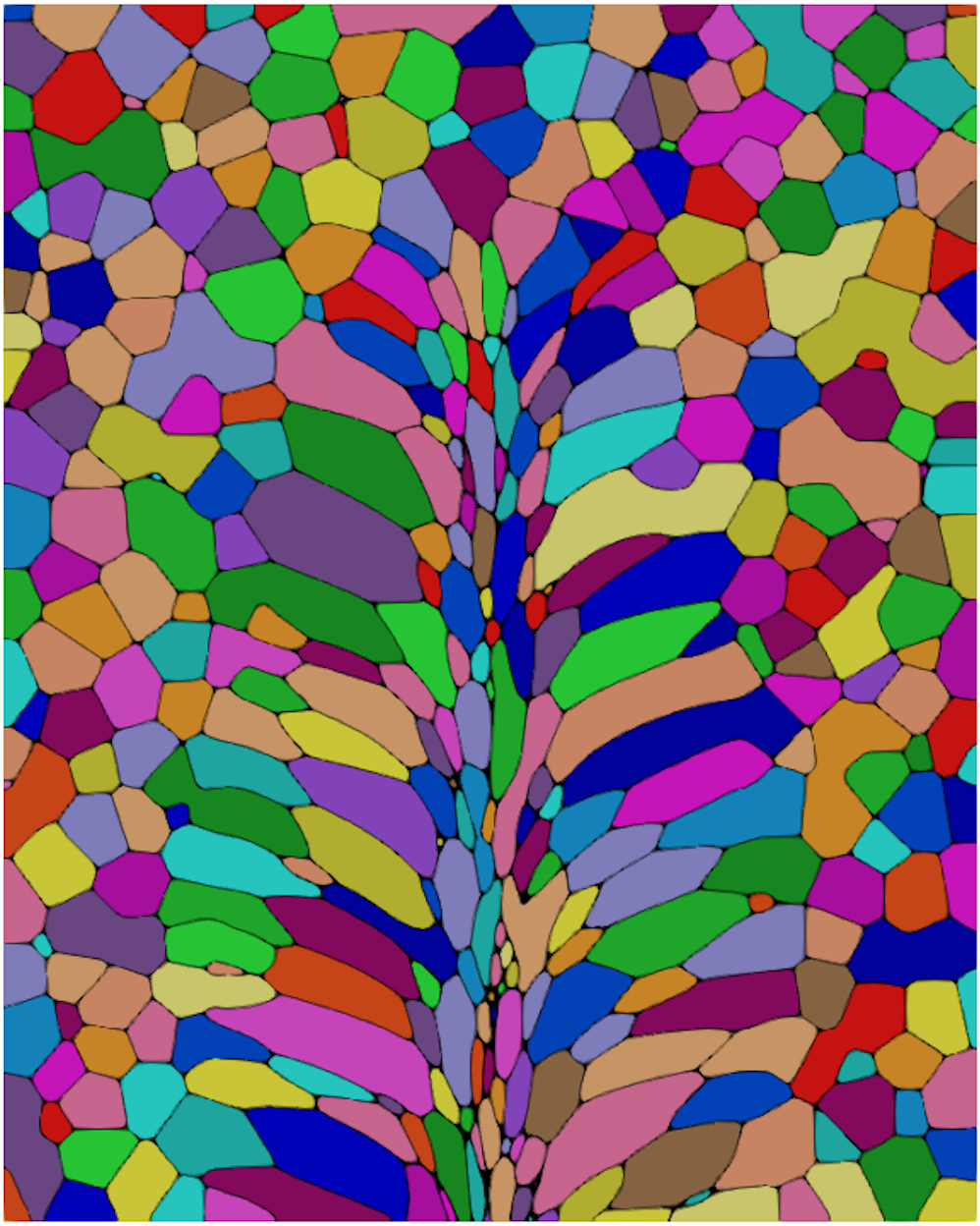}
  }
  \caption{ {Top surface of grain structures under different scan speeds with $\epsilon_4 = 0.11$}}
  \label{3d top vp}
\end{figure}

\begin{figure}[htbp]
  \centering
  \hspace{0\textwidth}
  \subcaptionbox{\,$V_p=1.0 $ m/s}[0.4\textwidth]{%
    \includegraphics[width=\linewidth]{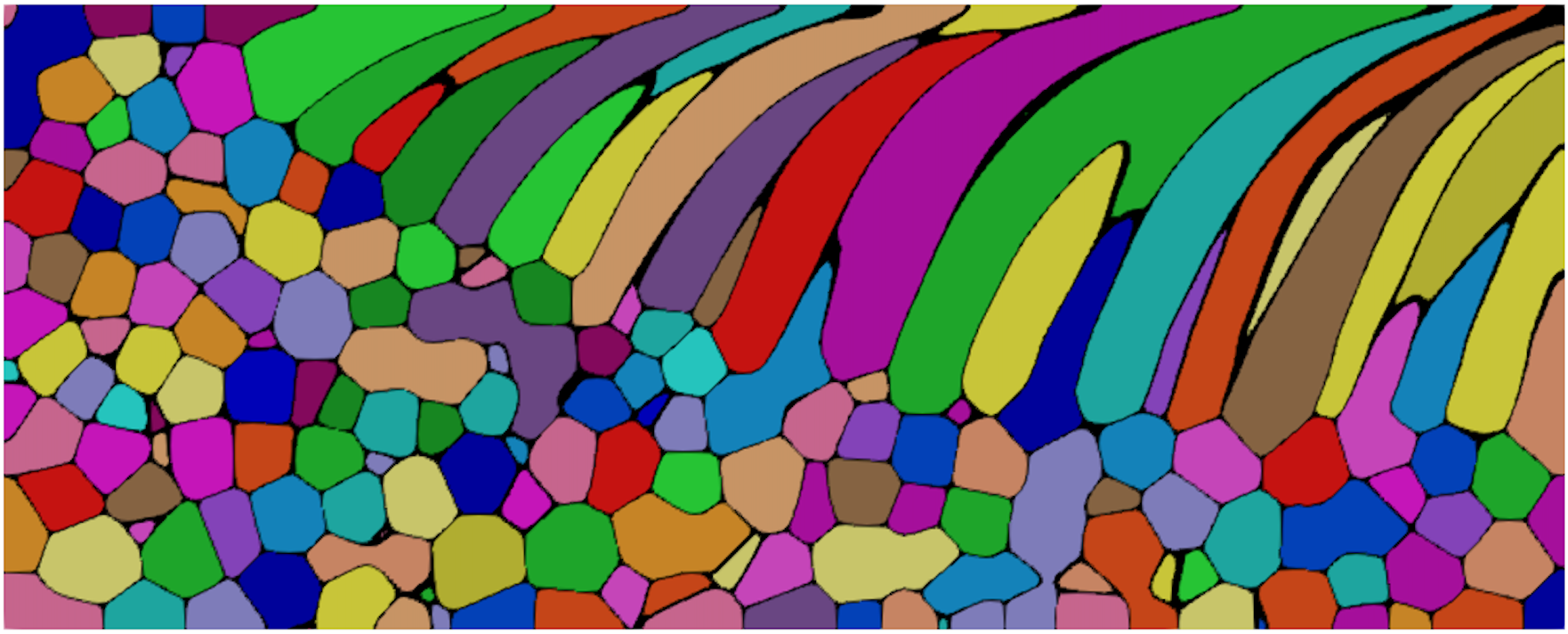}
  }
  \hspace{0.1\textwidth}
  \subcaptionbox{\,$V_p=1.5 $ m/s}[0.4\textwidth]{%
    \includegraphics[width=\linewidth]{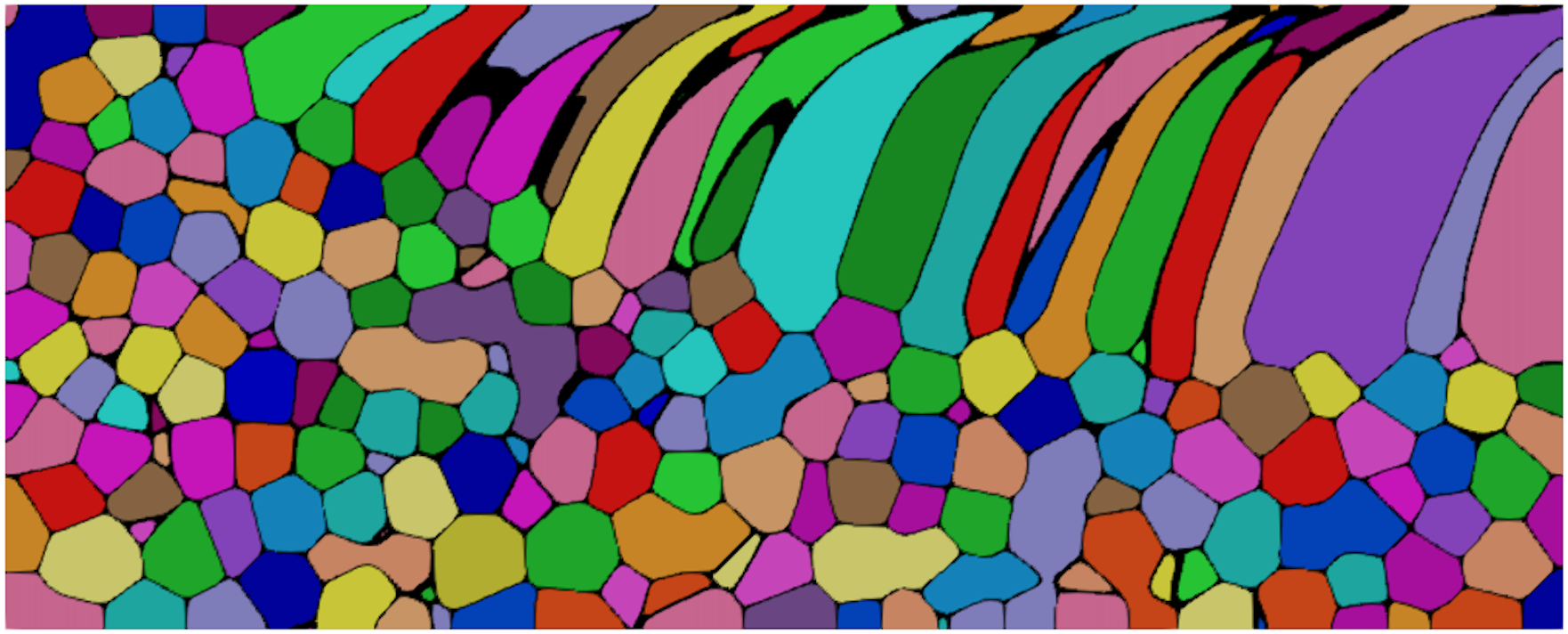}
  }
  \caption{ {Central cross section of grain structures under different scan speeds}}
  \label{3d cs vp}
\end{figure}

  {To study the effect of the laser scan speed}, we conducted two simulations with the scan speeds: $V_p=1.0 $ m/s and $V_p=1.5 $ m/s. The other parameters remain the same as in \tablename~\ref{tab:param},  {including the anisotropic parameter}. We illustrate  the top surface and the longitudinal central cross section of the simulated grain structures in \figurename~\ref{3d top vp} and \figurename~\ref{3d cs vp}, respectively.

By comparing these results of different scan speeds, we can see that increasing the scan speed globally reduces the depth and width of the melt pool,  {and a higher scan speed leads to a higher density of grains in the central melt pool and an increase of the grain curvature in the cross section. This  indicates that the grain selection effect was reduced under the high-speed condition, allowing more grains to grow. This might be due to the smaller melt pool size and the faster solidification, which limit the grain competition.} These trends are also consistent with the experimental and simulation results reported in \cite{Chadwick2025}.

Finally, we performed a statistical analysis for the 3D simulations with different scan speeds, using a geometric feature-based approach proposed in \cite{Chadwick2025, MacSleyne2008}. The idea is to characterize the morphology of the grains with three statistical metrics: the major, median, and minor axis lengths. Details on how to extract these information can be found in  { \ref{sec:App2}}.

\begin{figure}[htbp]
  \centering
  \hspace{0\textwidth}
  \subcaptionbox{\, Major axis length ($\upmu$m)}[0.49\textwidth]{%
    \includegraphics[width=\linewidth]{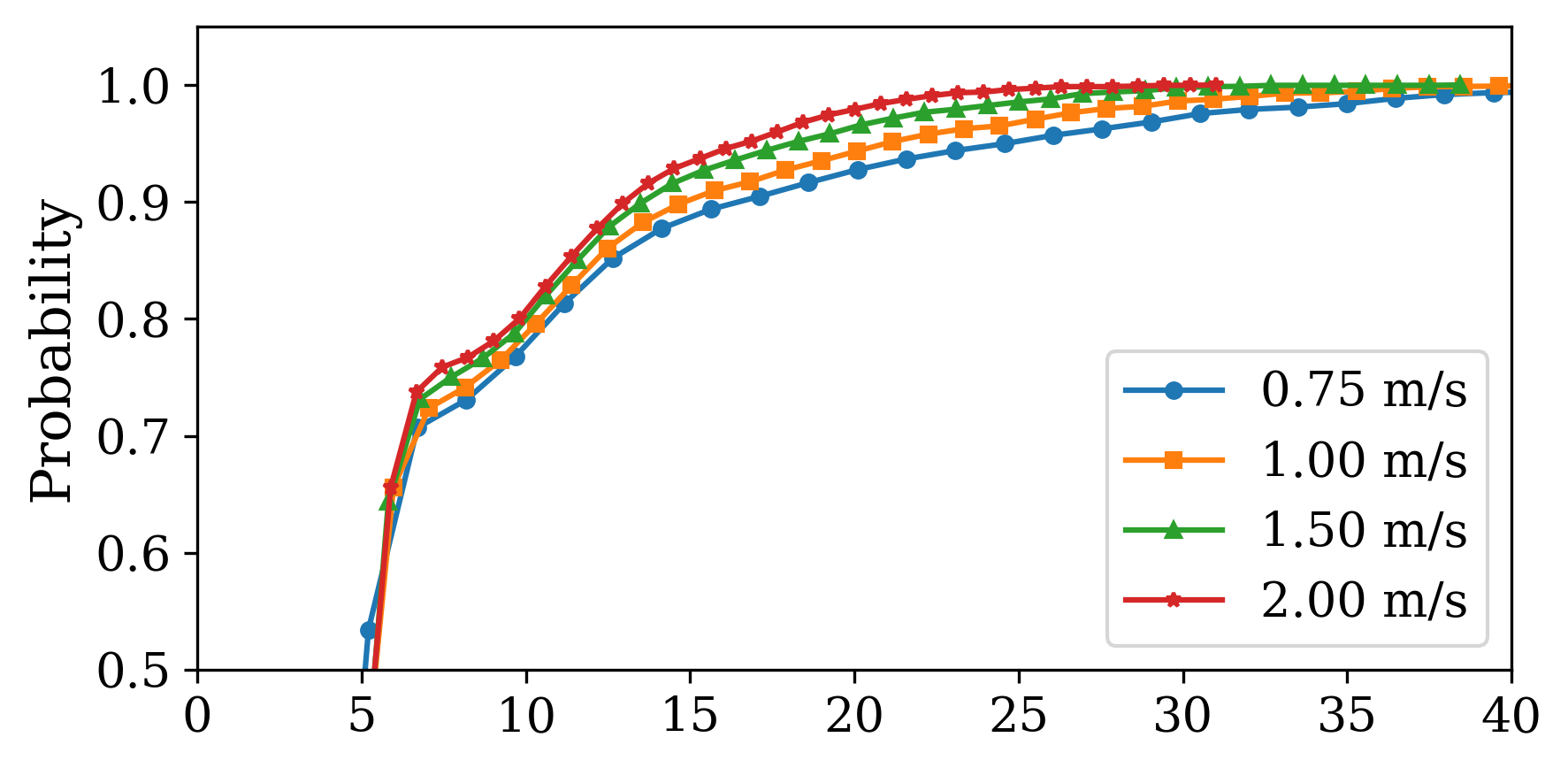}
  }
  \\
  \vspace{5pt}
  \hspace{0\textwidth}
  \subcaptionbox{\, Median axis length ($\upmu$m)}[0.49\textwidth]{%
    \includegraphics[width=\linewidth]{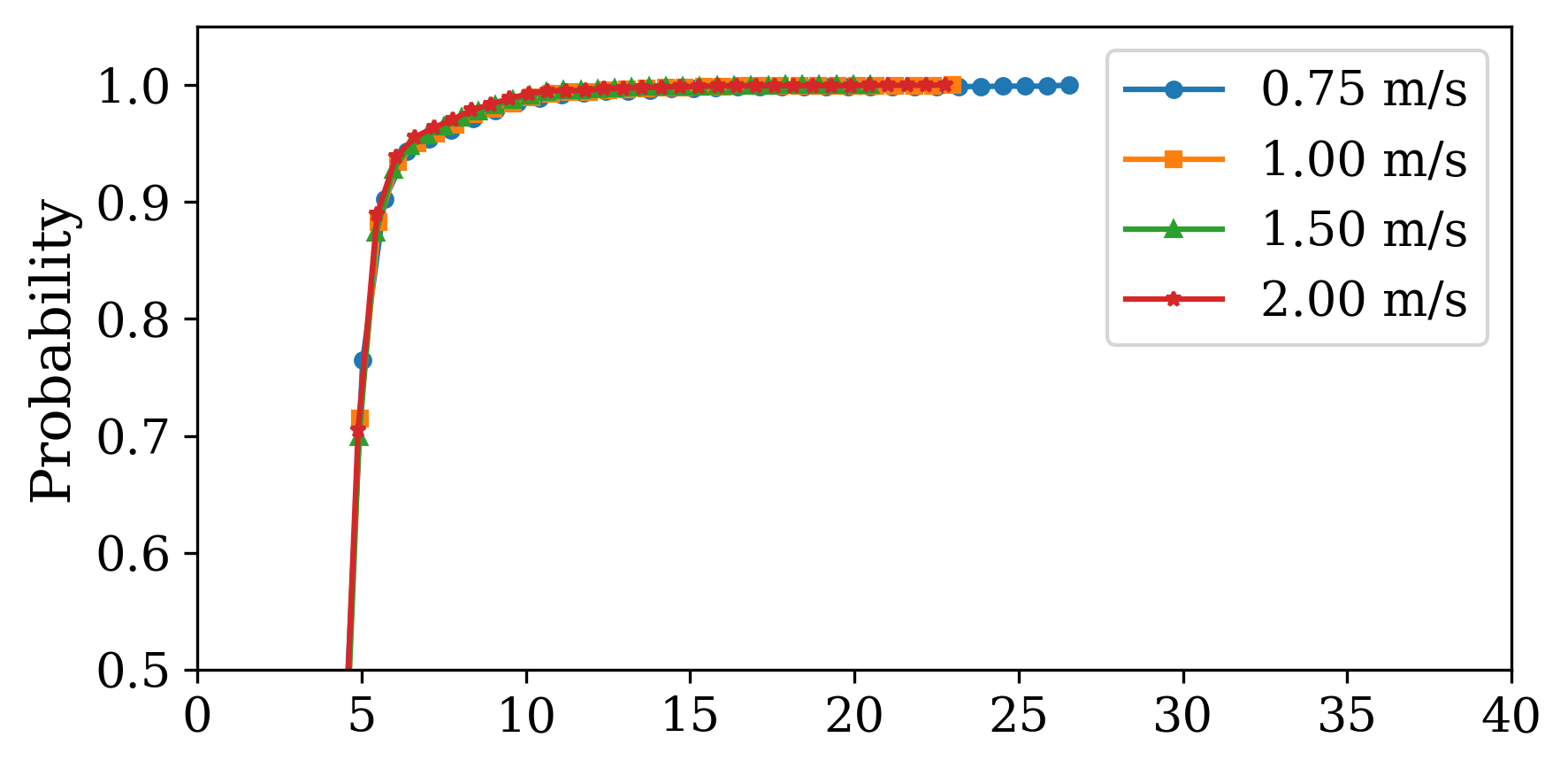}
  }
  \hspace{0\textwidth}
  \subcaptionbox{\, Minor axis length ($\upmu$m)}[0.49\textwidth]{%
    \includegraphics[width=\linewidth]{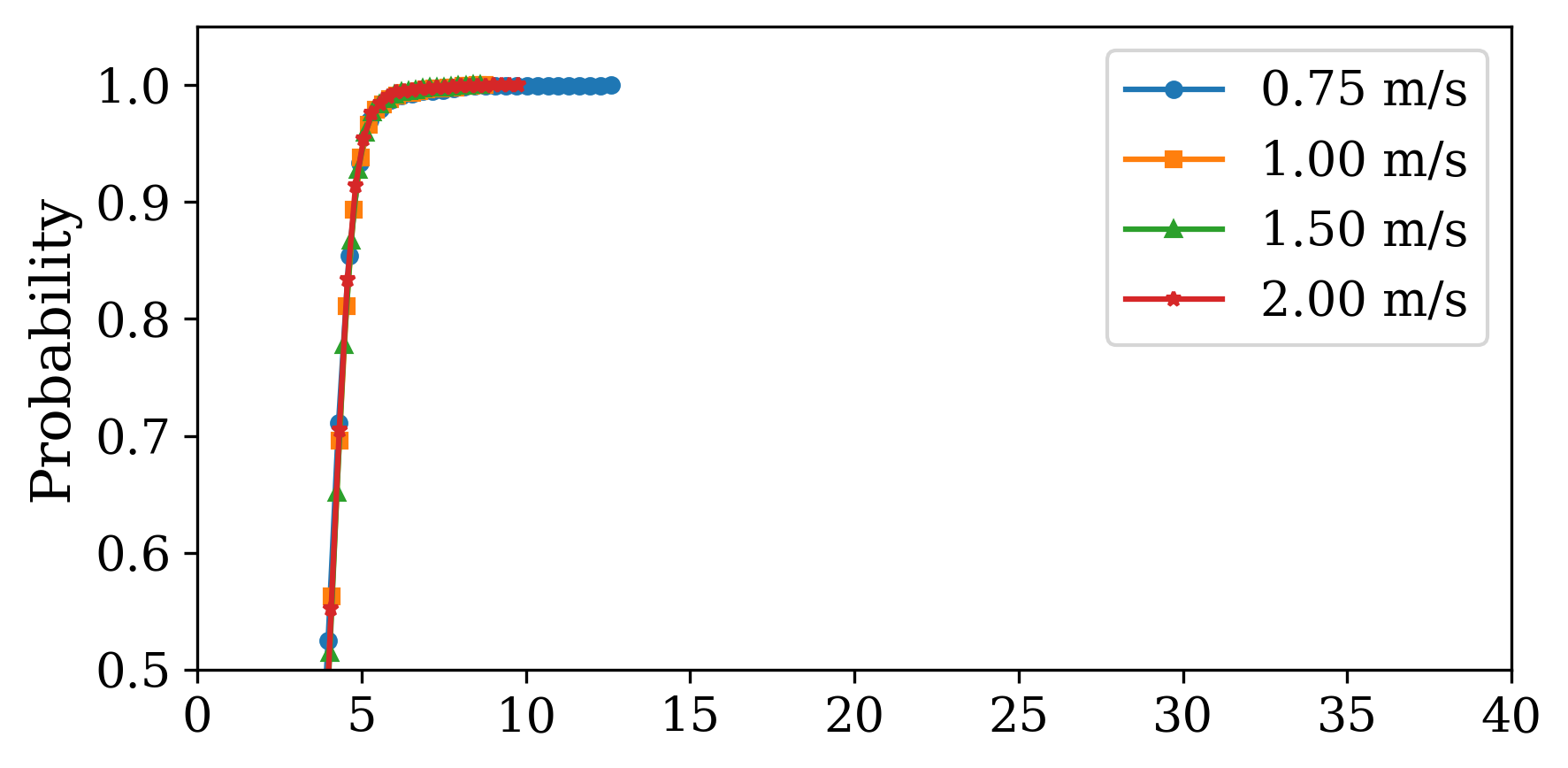}
  }
  \caption{ {Empirical cumulative distribution functions of the (a) major, (b) median, and (c) minor axis lengths for the simulated grain structures under different scan speeds}}
  \label{axis length}
\end{figure}

Figure~\ref{axis length} presents the empirical cumulative distribution functions of the major, intermediate, and minor axes across all grains, for selected laser scanning speeds: 0.75 m/s, 1.00 m/s, 1.50 m/s, and 2.00 m/s \cite{Berghaus2024}. We can see that only the major axis shows significant variations with different laser scan speeds. Among the selected conditions, the scan speed 0.75 m/s results in the longest average grain size, while 2.00 m/s produces the shortest. This trend aligns with the physical expectation that a larger melt pool allows for slower cooling and thus promotes greater grain growth. This confirms again our 3D simulations and the effectiveness of the proposed algorithm.

\FloatBarrier
\section{Conclusion}
This work proposed a finite element framework coupled with a class of stabilized semi-implicit time integration schemes for phase field simulations of grain growth during rapid solidification in AM processes. The key findings are summarized as follows.
\begin{enumerate}
    \item The developed stabilized semi-implicit schemes can enable  two orders-of-magnitude larger time steps than the conventional explicit scheme without sacrificing the accuracy, a speed-up of the same can be expected. 
    \item The proposed choice of the stabilization coefficient, along with the stability analysis, can ensure the revisited energy law, as confirmed by the numerical experiments.
    \item Both the first-order and second-order stabilized semi-implicit schemes showed  {superior} performance in terms of accuracy and stability.  {In general, the first-order scheme can serve as the default choice because of its simplicity of implementation. If the required accuracy cannot be achieved with a sufficiently large time step, the second-order scheme may instead be employed.}
    \item The proposed framework can accurately capture detailed microstructure evolution, including the  grain morphology and growth characteristics observed in rapid solidification during AM processes.  
    \item The simulations and numerical results are validated by investigating the effects of kinetic anisotropy and different laser scan speeds and are consistent with that reported in  the literature \cite{Chadwick2021}.
\end{enumerate}

Our ongoing work is to incorporate convolution tensor decomposition \cite{Lu2025} to further accelerate large volume phase field simulations.  Additionally, adaptive time-stepping strategies can be developed within the proposed framework. These advancements will enable efficient high fidelity microstructure simulations and ultimately support the improvement of 3D printing processes. Furthermore, other effects, e.g., composition variations and energy anisotropy, can be included in the same  framework for different alloys and solidification conditions.

\section*{Acknowledgements}
CY and YL would like to acknowledge the support of University of Maryland Baltimore County through the startup fund and the COEIT Interdisciplinary Project Award. {This work was carried out using the High Performance Computing Facility (HPCF) at University of Maryland Baltimore County.} 

\appendix
\section{Input temperature field}
\label{sec:App1}
For the temperature profile, we follow the settings in \cite{Chadwick2021}, which reads
\begin{equation}
T(X, R) = T_0 + \frac{Q}{2\pi \kappa_T}\left(\frac{1}{R} \right)\text{exp}\left(-\frac{V_p}{2c} (R+X) \right),
\end{equation}
where $T_0$ denotes the environment temperature, $Q$ is the absorbed heat power from the source, $\kappa_T$ is the thermal conductivity, $c$ is the thermal diffusivity, and $V_P$ is the velocity of the moving heat source. 

In this equation, $R = \sqrt{X^2+Y^2+Z^2}$ represents the radial distance from the heat source, where $X$, $Y$, and $Z$ are the spatial coordinates in the moving reference frame. And it can be related to the simulation frame with $X = x - x_0 - V_p * t$, $Y=y$ and $Z=z-z_0$, where x, y, and z are the simulation coordinates and $x_0$, $z_0$ are initial coordinates of the heat source. The parameters of Rosenthal's solution are given in \tablename~\ref{tab:thermal}. It has been shown that the Rosenthal model provides  cooling rate predictions that align closely with experimental measurements and more advanced thermal modeling approaches for  the powder bed fusion of 316L stainless steel \cite{ScipioniBertoli2019}.

\begin{table}[h]
\centering
\caption{The parameters for the Rosenthal's solution}
\begin{tabular}{lll}
\toprule
\textbf{Parameter} & \textbf{Value} & \textbf{Reference} \\
\midrule
$\kappa_T$ & $2.7 \times 10^{-5}\ \text{W}. \,\upmu\text{m. K}$  & \cite{Kim1975} \\
$c$ & $5.2 \times 10^6\ \upmu\text{m}^2/\text{s}$ & \cite{Kim1975} \\
$Q$ & $25\ \text{W}$ & \\
$V_P$ & $10^6\ \upmu\text{m}/\text{s}$ & \\
$T_0$ & $300\ \text{K}$ & \\
$x_0$ & $50\ \upmu\text{m}\ \text{in 2D}, \, 120\ \upmu\text{m}\ \text{in 3D}$ & \\
$z_0$ & $21.2 \ \upmu \text{m}$ & \\
\bottomrule
\end{tabular}
\label{tab:thermal}
\end{table}

\section{Statistical analysis of  grains}
\label{sec:App2}
To extract individual grains, we first construct a directed graph of the grain network in the sparse matrix form, then apply the 'connected components'  from the Scipy package to identify the strongly connected components and thereby segment the individual grains \cite{Pearce2005}. For each grain, we define an indicator function $D(r)$ such that
\begin{equation}
D(r) =
\begin{cases}
    1, & r\text{ inside a specific grain} \\
    0, & r\text{ outside a specific grain}.
\end{cases}
\end{equation}
Using this definition, we compute the 3D central moments with respect to the grain’s centroid $(x_n,y_n,z_n)$ as 
\begin{equation}
    \upsilon_{ijk} = \int_{V} (x-x_n)^i(y-y_n)^j(z-z_n)^k D(r) \, dV,
\end{equation}
where $V$ is the volume occupied by the grain. The second-order central moments form a symmetric tensor
\begin{equation}
\mathcal{I} = 
\begin{pmatrix}
\upsilon_{020} + \upsilon_{002} & -\upsilon_{110} & -\upsilon_{101} \\
-\upsilon_{110} & \upsilon_{200} + \upsilon_{002} & -\upsilon_{011} \\
-\upsilon_{101} & -\upsilon_{011} & \upsilon_{200} + \upsilon_{020}
\end{pmatrix},
\end{equation}
which resembles an inertia tensor and characterizes the spatial distribution of the grain mass. We then perform eigen-decomposition of the central moment tensor $\mathcal{I}$, i.e., $\mathcal{I} = \boldsymbol{{}\Upsilon A \Upsilon^{-1}}$, where $\boldsymbol{\Upsilon}$ is an orthogonal matrix whose columns represent the principal axes, and $\boldsymbol{A}$ is a diagonal matrix representing the moment tensor in the principal axis coordinate system, which reads
\begin{equation}
\boldsymbol{A} = 
\begin{pmatrix}
\bar{\upsilon}_{020} + \bar{\upsilon}_{002} & 0 & 0 \\
0 & \bar{\upsilon}_{200} + \bar{\upsilon}_{002} & 0 \\
0 & 0 & \bar{\upsilon}_{200} + \bar{\upsilon}_{020}
\end{pmatrix},
\end{equation}
where $\bar{\upsilon}_{ijk}$ is the central moments in the principal axis coordinates. We then calculate the $\bar{\upsilon}_{ijk}$ using the eigenvalues
\begin{align}
\bar{\upsilon}_{200} &= \frac{1}{2}(\lambda_2 + \lambda_3 - \lambda_1) \notag \\
\bar{\upsilon}_{020} &= \frac{1}{2}(\lambda_1 + \lambda_3 - \lambda_2) \notag \\
\bar{\upsilon}_{002} &= \frac{1}{2}(\lambda_1 + \lambda_2 - \lambda_3),
\end{align}
where $\lambda_1, \lambda_2, \lambda_3$ are the eigenvalues of the matrix $\mathcal{I}$. 
Based on the computed central moments, the semi-axes lengths of the equivalent ellipsoid are given by $a = (5\bar{\upsilon}_{200}/V)^{0.5}$, $b = (5\bar{\upsilon}_{020}/V)^{0.5}$ and $c = (5\bar{\upsilon}_{002}/V)^{0.5}$, where $V$ is the volume of the grain. These values are then doubled to obtain the full lengths of the major, median, and minor axes of the grain.

\bibliographystyle{main.bst} 
\bibliography{main.bib}






\end{document}